\documentclass[pra,aps,twocolumn]{revtex4}
\usepackage[12pt]{xy}
\usepackage{graphicx}
\usepackage{colordvi}
\usepackage{color}

\begin{document}

\title{Wigner Function Evolution of Quantum States in Presence of
  Self-Kerr Interaction}

\author{Magdalena Stobi\'nska} \email{magda.stobinska@fuw.edu.pl}
\affiliation{Institute f\"ur Optik, Information und Photonik,
  Max-Planck Forschungsgruppe, Universit\"at
  Erlangen-N\"urnberg,\\ G\"unter-Scharowsky-Str.\ 1, Bau 24, 91058,
  Germany}

\author{G. J. Milburn}
\affiliation{Centre for Quantum Computer Technology and School of
  Physical Sciences, The University of Queensland, St Lucia,
  Queensland 4072, Australia}

\author{Krzysztof W\'odkiewicz}
\email{wodkiew@fuw.edu.pl} \affiliation{Instytut Fizyki Teoretycznej,
  Uniwersytet Warszawski, Warszawa 00--681, Poland}

\affiliation{Department of Physics and Astronomy, University of New
  Mexico, Albuquerque, NM~87131, USA} \date{\today}

\begin{abstract}

  A Fokker-Planck equation for the Wigner function evolution in a
  noisy Kerr medium ($\chi^{(3)}$ non-linearity) is presented.  We
  numerically solved this equation taking a coherent state as an
  initial condition.  The dissipation effects are discussed. We
  provide examples of quantum interference, sub-Planck phase space
  structures, and Gaussian versus non-Gaussian dynamical evolution of
  the state.  The results also apply to the description of a
  nanomechanical resonator with an intrinsic Duffing nonlinearity.

\end{abstract}

\maketitle

\section{Introduction}

Nonlinear interaction of light in a medium provides a very useful
framework to study various nonclassical properties of quantum states
of radiation.  These nonclassical properties are usually associated
with quantum interference and entanglement.  The phase space Wigner
distribution function description of quantum states of light is a
powerful tool to investigate such nonclassical effects. With the help
of the Wigner function one can simply visualize quantum
interference. For example, a signature of quantum interference is
exhibited in the Wigner function by the non-positive values and
sub-Planck structures \cite{Ludmila}.  The non-positive Wigner
function is a witness of a nonclassicality and monitors a decoherence
process of a quantum state, e.g.\ a photon-added coherent state in the
photon-loss channel \cite{Shang}, photon-subtracted squeezed state
\cite{Agarwal}, Gaussian quadrature-entangled single photon subtracted
light pulse \cite{Grangier2007}, two-photon Fock state
\cite{Grangier2006}, odd photon number states superposition
\cite{Polzik} and coherent states superpositions \cite{Jeong}. The
Wigner function has also been used for computing numerically the
quantum-mechanical corrections to the classical dynamics of a
nanomechanical resonator (its characteristic pattern, the interference
fringes and negative values, served as a signature for a classical to
quantum domain transition) \cite{Katz}; it has been applied in the
model of the dynamics of a nanoscale semiconductor laser
\cite{Weetman}, to list only a few examples of its applications.

The Kerr medium provides one of the simplest nonlinearity for which
there exists a simple analytic Wigner function \cite{Wigner}
expression.  The highly $\chi^{(3)}$ nonlinear systems generated a lot
of interest recently due to their applications e.g.\ to nondemolition
measurements~\cite{Imoto}, quantum computing
architectures~\cite{Turchette}, single particle
detectors~\cite{Mohapatra}.  The well known example of such a medium
is an optical fiber.  However, the nonlinearity is small in fibers and
is often accompanied by other unwanted effects.  Enhanced Kerr
nonlinearity was studied in terms of electromagnetically induced
transparency~\cite{Imamoglu} and was observed in Bose Einstein
condensates~\cite{Hau} and cold atoms~\cite{Kang}. Recent proposals
predict obtaining enormous Kerr nonlinearity using the Purcell
effect~\cite{Bermel}, Rydberg atoms~\cite{Mohapatra}, and interaction
of a cavity mode with atoms~\cite{Brandao}. The first and last method
is predicted to obtain the nonlinearity of nine orders of magnitude
higher than natural self-Kerr interactions with negligible losses. The
significant nonlinearity has also been observed for nanomechanical
resonators~\cite{Kozinsky}.

Most of the investigations of the Wigner function of light have been
made for steady state situations. The simplicity of the Kerr medium
will allow to study the full time-dependent Wigner function dynamics
with or without a quantum noise.

In this paper we present a Fokker-Planck equation which determines the
time evolution of the Wigner function in a noisy $\chi^{(3)}$ medium.
We solve this equation numerically assuming a coherent state as an
initial condition. We discuss first an ideal and then a dissipative
Kerr medium. The results obtained for the ideal case reveal the
quantum nature of the state under evolution. The coherent state, known
as the most classical among all the pure states, becomes a
non-Gaussian squeezed state after some time of interaction with the
medium.  For some specific times it becomes a finite superposition of
other coherent states. An interference pattern with the negative
values is clearly visible on the plots of its Wigner function.  Due to
the small value of the $\chi^{(3)}$ nonlinearity and losses in optical
fibers such phenomena have never been observed for light; neither for
any other system.  However, it turns out that not all quantum effects
are washed out due to the decoherence. The Fokker-Planck equation
allows for a similar state evolution analysis for any other input
state.

The paper is organized as follows: in section \ref{fs} the
Fokker-Planck equation for the Wigner function evolution in a
$\chi^{(3)}$ medium is derived from the Master equation obtained for a
single mode of light density operator. The equation is displayed in a
polar coordinates. The decoherence effects: losses and thermal noise
are included. The initial and boundary conditions for the Wigner
function are set.  In section \ref{ss} the numerical results of the
Wigner function evolution in a nondissipative $\chi^{(3)}$ medium are
introduced and discussed. These results are obtained using three
independent methods: computing the Fokker-Planck equation and the
other two equations determining the Wigner function directly, which
are obtained from its definition.  Correspondence between the Wigner
function negative values and zeros of the Q-function is noted.  In
section \ref{ts} the influence of the decoherence process on quantum
effects such as the interference pattern and the negative values of
the Wigner function is analyzed. The technical limitations on the use
of the Fokker-Planck equation are discussed. The Wigner function
sub-Planck structure is shown in section \ref{sub}. In section
\ref{fos} the numerical methods used in sections \ref{ss} and \ref{ts}
are compared and discussed.  Finally some conclusions are presented.

\section{The Fokker-Planck Equation for a Self-Kerr Interaction}\label{fs}

The interaction Hamiltonian for a totally degenerate four-wave mixing
process, e.g. in an optical fiber, is of second-order in creation
$a^{\dagger}$ and annihilation $a$ light operators
\begin{equation}
H = \hbar \frac{\kappa}{2}\, a^{\dagger}a^{\dagger}a\, a,
\end{equation}
where $\kappa$ is a nonlinear constant proportional to
$\chi^{(3)}$~\cite{Tanas}.  Please note that similar Hamiltonian
describes a single nanomechanical resonator with $a^{\dagger}$ and $a$
being raising and lowering operators related to its position and
momentum operators, and $\kappa$ proportional to the Duffing
nonlinearity~\cite{Woolley}.

In a general case, including damping and thermal noise in the medium,
a one-mode density operator evolution, both for light and for a
nanomechanical resonator, is determined by the Master
equation~\cite{Walls}
\begin{eqnarray}
\partial_t \rho(\tau) &=& - \frac{i}{\hbar}[H,\rho(\tau)] +
\frac{\Gamma}{2}\left([a\rho(\tau),a^{\dagger}] + [a, \rho(\tau)
  a^{\dagger}]\right)
\nonumber\\
&+& \Gamma N [[a,\rho(\tau)],a^{\dagger}],
\label{Kerr_Master}
\end{eqnarray}
\noindent
where $\tau = -\kappa t$ is a unitless evolution parameter, $t$ is the
interaction time, $\Gamma$ is a damping constant, $N =
1/\{\exp\left(\frac{\hbar \omega}{kT}\right)-1\}$ is a mean number of
photons in a thermal reservoir.

The solutions of the above equation are well known~\cite{Tanas,
  Milburn1986, Milburn1989, Perinova1990} but since it is an
operator-valued equation, they are inappropriate for computer
simulations.  A Fokker-Planck type evolution equation for the Wigner
function can be easily obtained from equation~(\ref{Kerr_Master})
using the standard quantum optics phase space
methods~\cite{Gardiner}. Since every density operator determines its
Wigner function uniquely, the knowledge of its evolution is equivalent
to the full knowledge of the density operator dynamics.

The dynamics of the Wigner function in a dissipative medium with a
self-Kerr interaction is governed by the Fokker-Planck equation, which
takes the following form in the polar coordinates $\gamma =
re^{i\varphi}$
\begingroup \setlength{\arraycolsep}{0cm}
\begin{eqnarray}
\partial_{\tau} W(\tau, r, \varphi) &=&
\left\{ (r^2-1)\partial_{\varphi} - \frac{1}{16}
    \left(\frac{1}{r}\partial_r\partial_{\varphi} + \partial^2_r
      \partial_{\varphi} + \frac{1}{r^2}\partial^3_{\varphi} \right) \right.
\nonumber\\
&+& \left. \xi + \frac{\xi}{2} \left(r +
    \frac{1}{2}\left(\frac{1}{2} +
       N\right) \frac{1}{r}\right) \partial_r \right.
\nonumber\\
&+& \left.
   \frac{\xi}{4}\left(\frac{1}{2} +
       N\right) \left(\partial^2_r +
      \frac{1}{r^2}\partial^2_{\varphi}\right) \right\}
W(\tau, r, \varphi),
\label{eq:Fokker-Planck}
\end{eqnarray}
\endgroup
where $\gamma$ is a point in a phase space, $\xi = \Gamma/\kappa$.

This is a third-order nonlinear differential equation. We compute this
equation for the following initial and boundary conditions.  The
evolution starts with a coherent state $|\alpha\rangle$ described by a
Wigner function $W(0, \gamma) = \frac{2}{\pi}
e^{-2|\alpha-\gamma|^2}$. The Wigner function tends to zero in the
infinity $\lim_{r \to \infty} W(\tau,r,\varphi) = 0$.  Since some of
the coefficients in equation~(\ref{eq:Fokker-Planck}) are singular for
$\gamma=0$, we take $W(\tau,0,0) =
\frac{2}{\pi}e^{-2|\alpha|^2e^{-\tau \xi}}$, which was derived from
the Master equation.

\section{The nonlinear part of evolution}\label{ss}

For $\xi=0$ and $N=0$ the Fokker-Planck
equation~(\ref{eq:Fokker-Planck}) reduces to its first line and
describes the evolution in an ideal Kerr medium.

\begin{figure*}[h!]
\begin{center}
  \scalebox{0.5}{\includegraphics{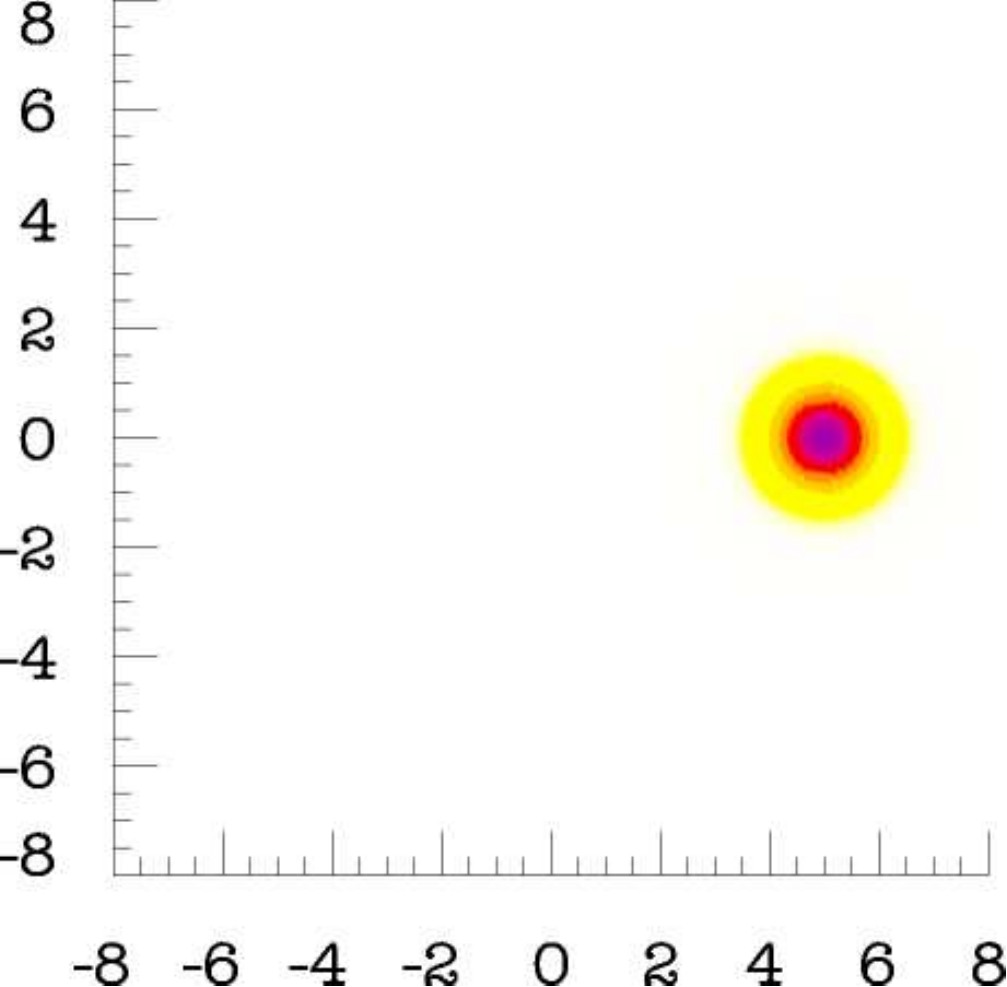}}
  \scalebox{0.5}{\includegraphics{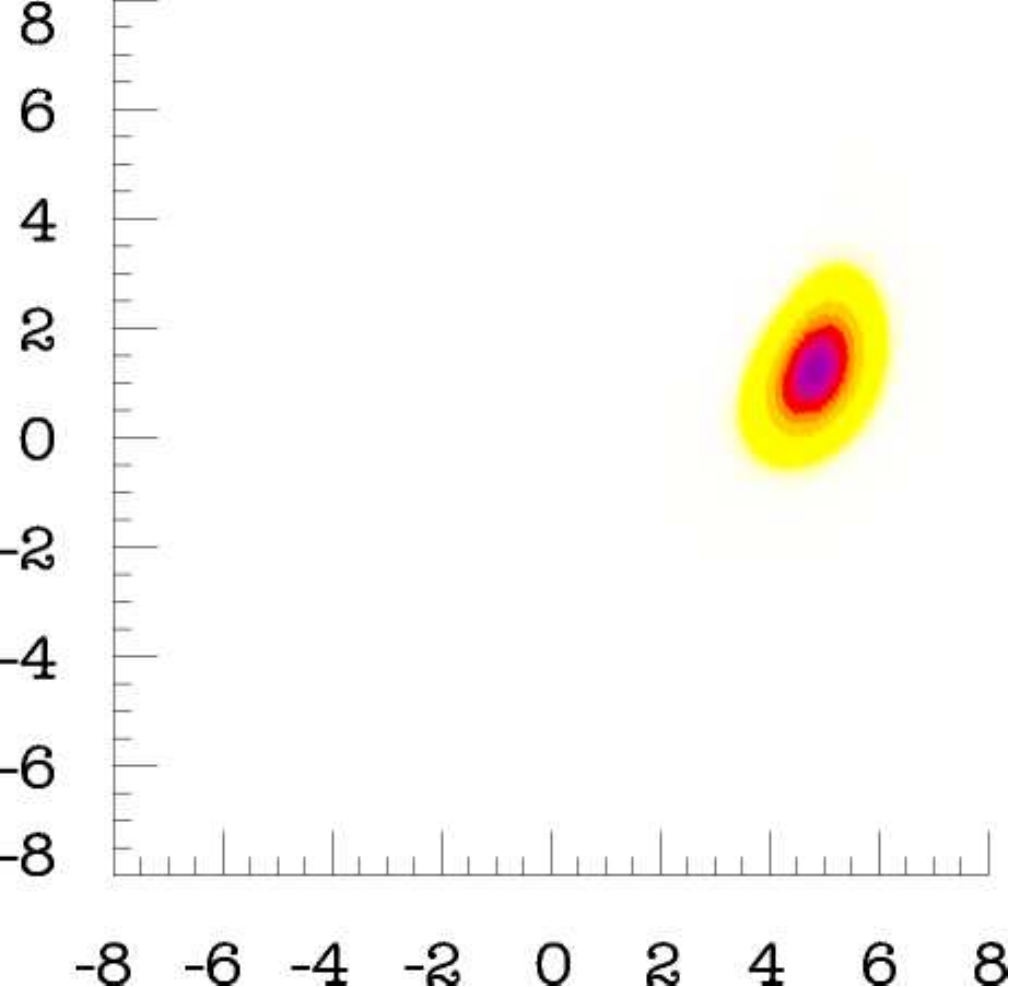}}
  \scalebox{0.5}{\includegraphics{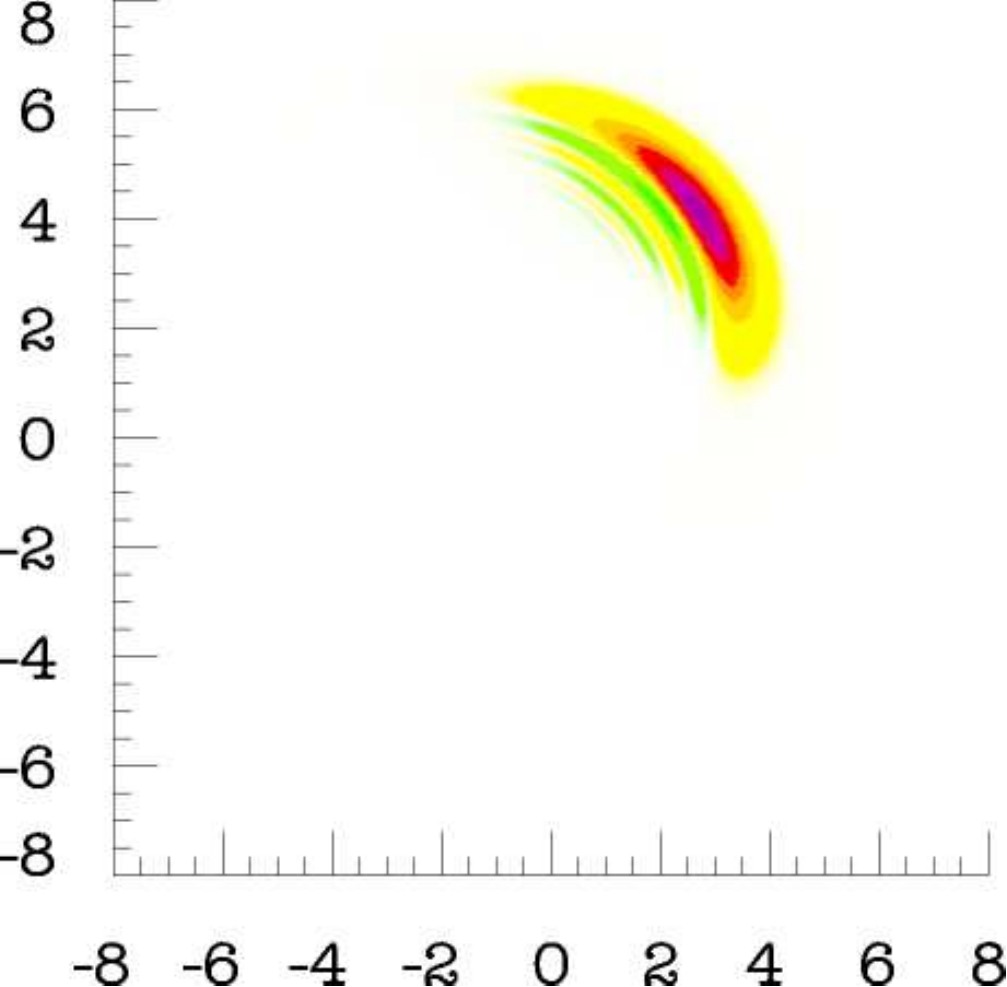}}
  \scalebox{0.5}{\includegraphics{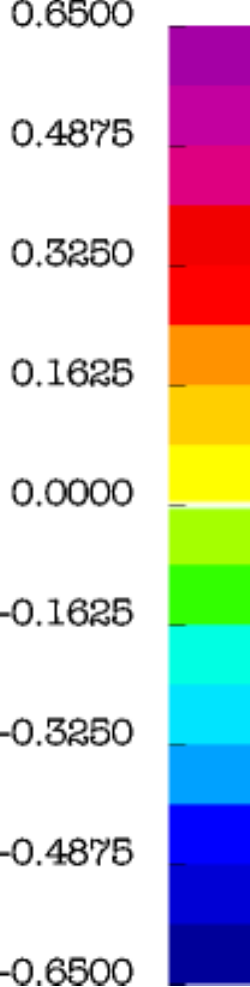}}
\end{center}
\caption{(Color online) The Wigner function for $\alpha=5$ and $\tau=0$ is a Gaussian
  function for the coherent state -- the left figure. Due to the fact
  that the evolution in an ideal $\chi^{(3)}$ medium is periodic, we
  achieve the same shape of the Wigner function for $\tau=2\pi$. The
  distribution is an ellipse and the state becomes squeezed beginning
  from $\tau=0.01$ -- the middle figure. The negative values of the
  function start to appear for $\tau = 0.04$ -- the right figure.}
\label{Fig:1}
\end{figure*}

\begin{figure*}[h!]
\begin{center}
  \scalebox{0.5}{\includegraphics{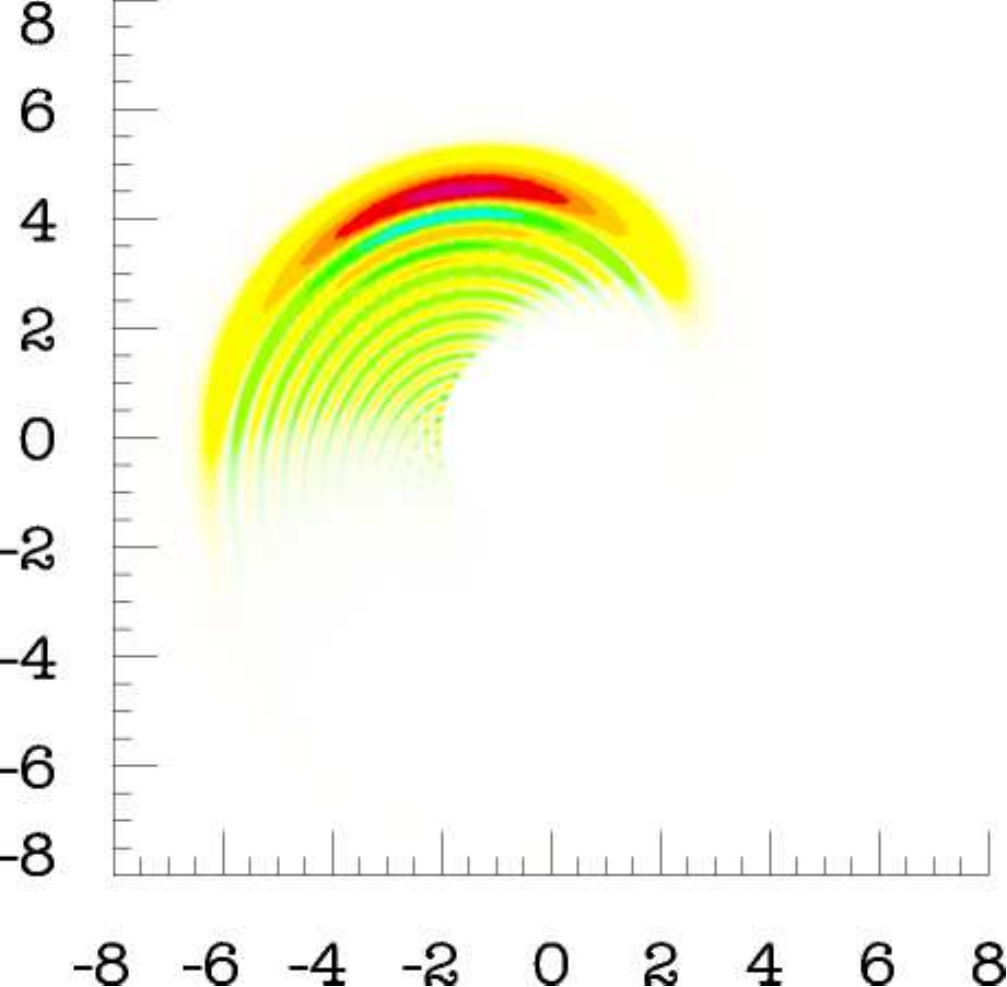}}
  \scalebox{0.5}{\includegraphics{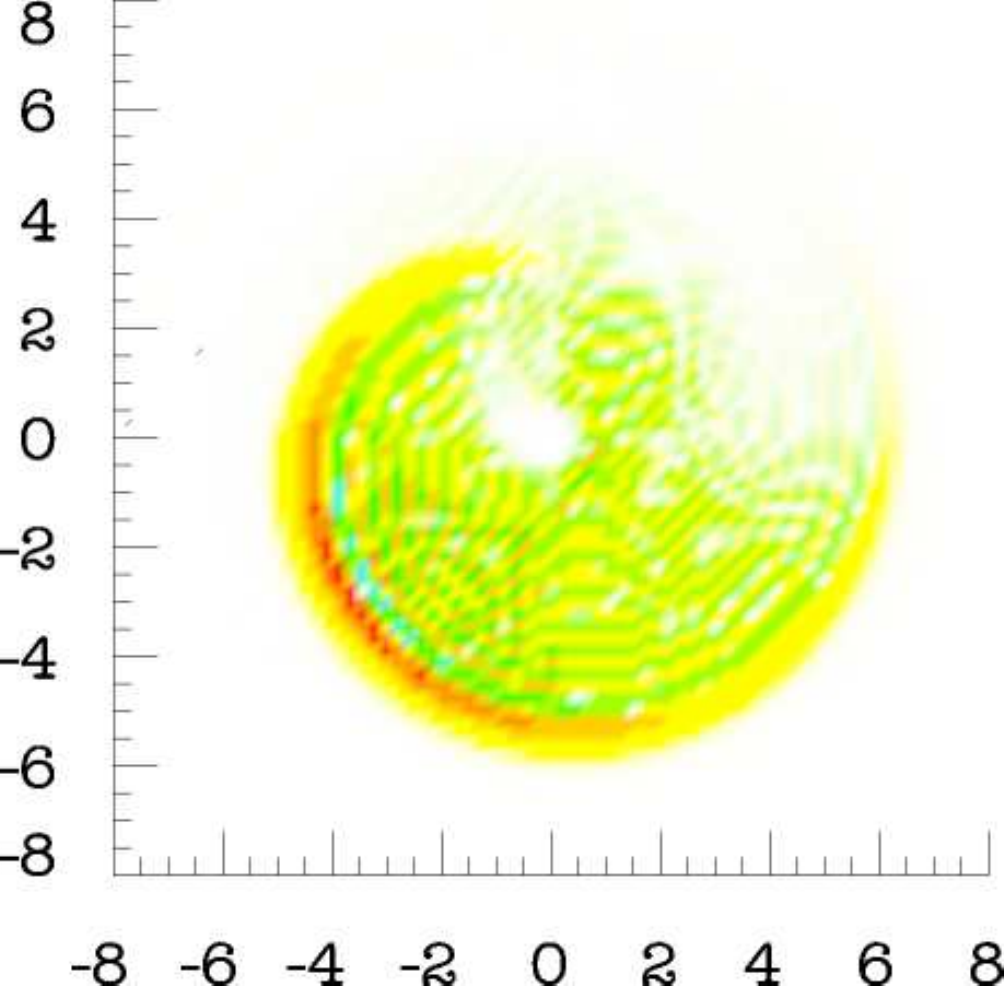}}
  \scalebox{0.5}{\includegraphics{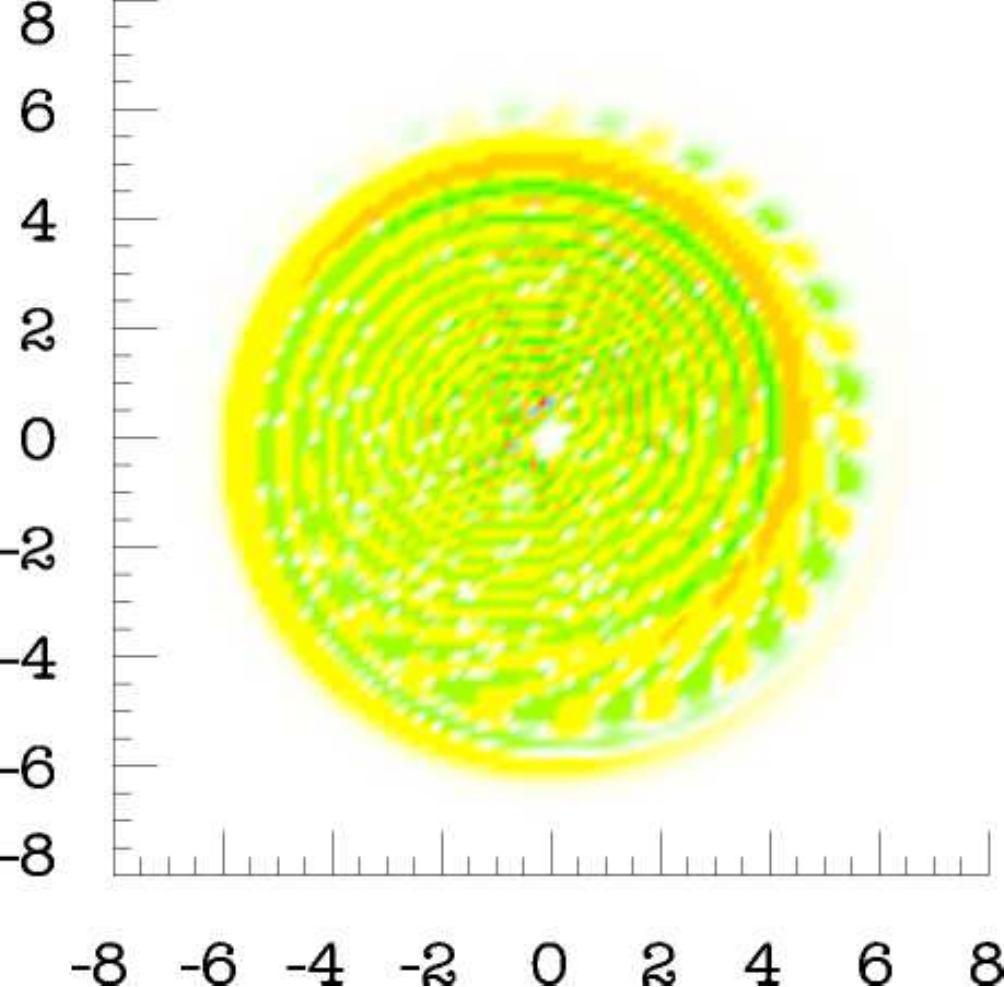}}
  \scalebox{0.5}{\includegraphics{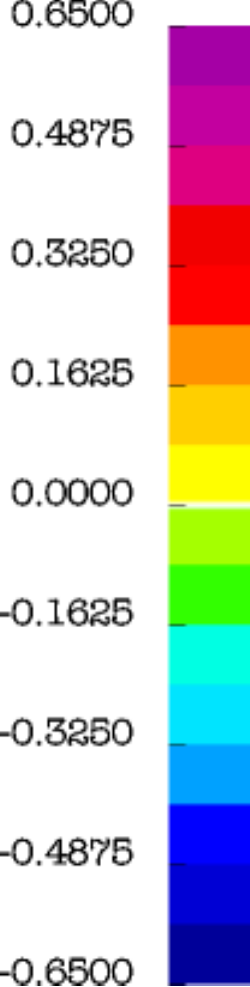}}
\end{center}
\caption{(Color online) Further steps of Wigner function evolution obtained also for
  $\alpha=5$ and $\tau = 0.08$ -- the left figure, $\tau=0.16$ -- the
  middle figure and $\tau = 0.3$ -- the right figure.  An interference
  pattern arises in the form of a ``tail'' of interference fringes.}
\label{Fig:2}
\end{figure*}

\begin{figure*}[h!]
\begin{center}
  \scalebox{0.5}{\includegraphics{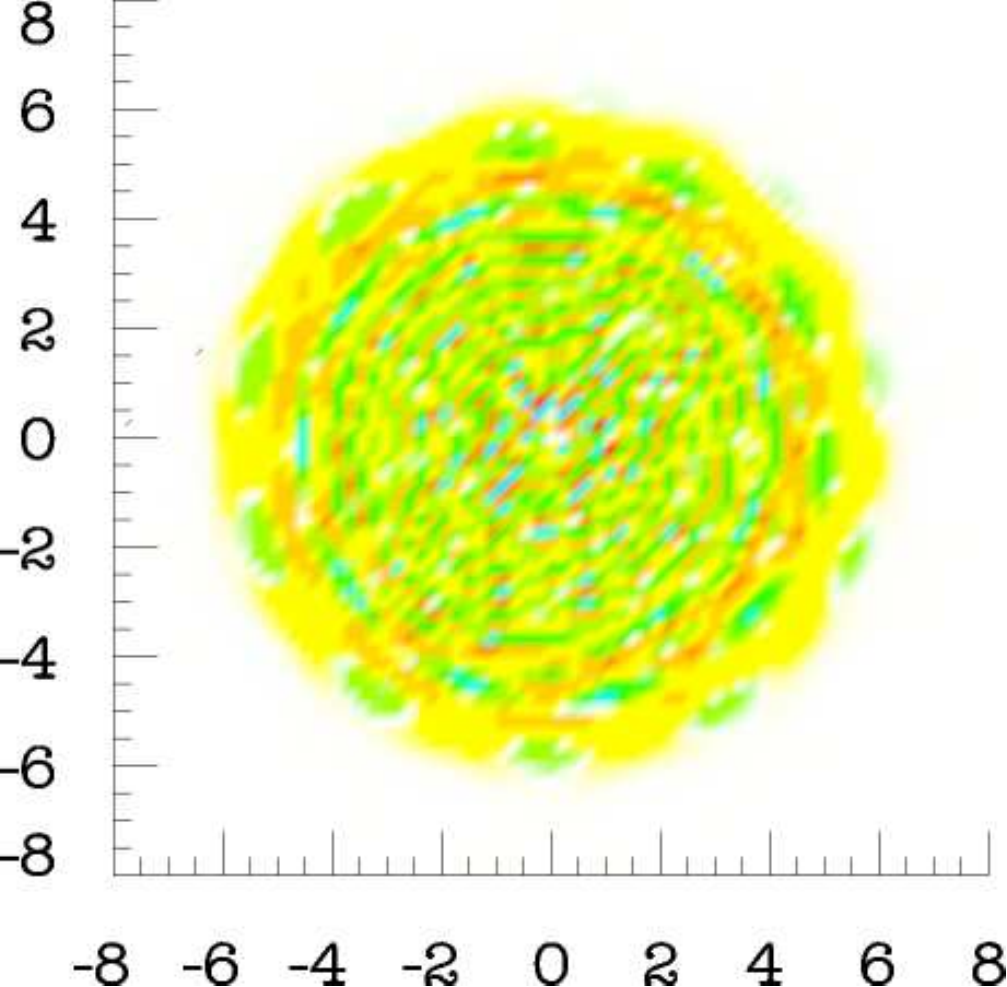}}
  \scalebox{0.5}{\includegraphics{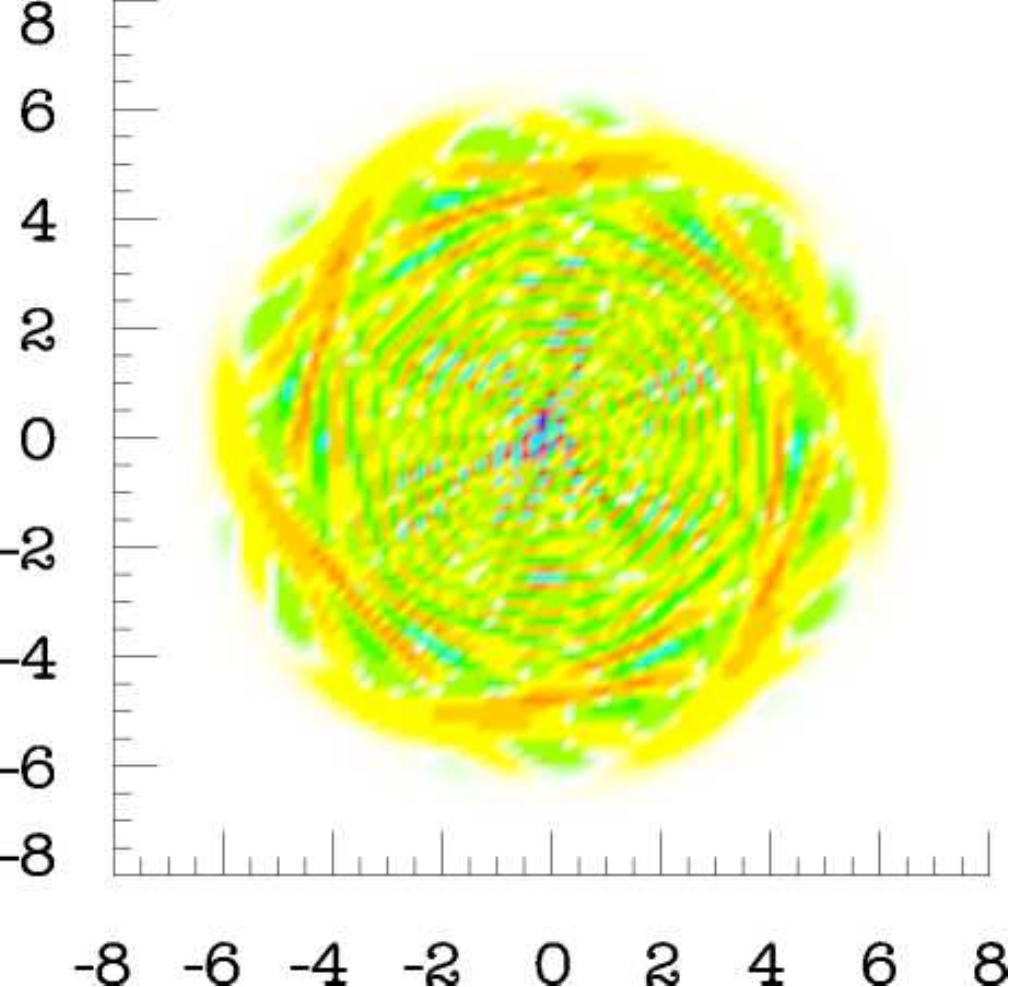}}
  \scalebox{0.5}{\includegraphics{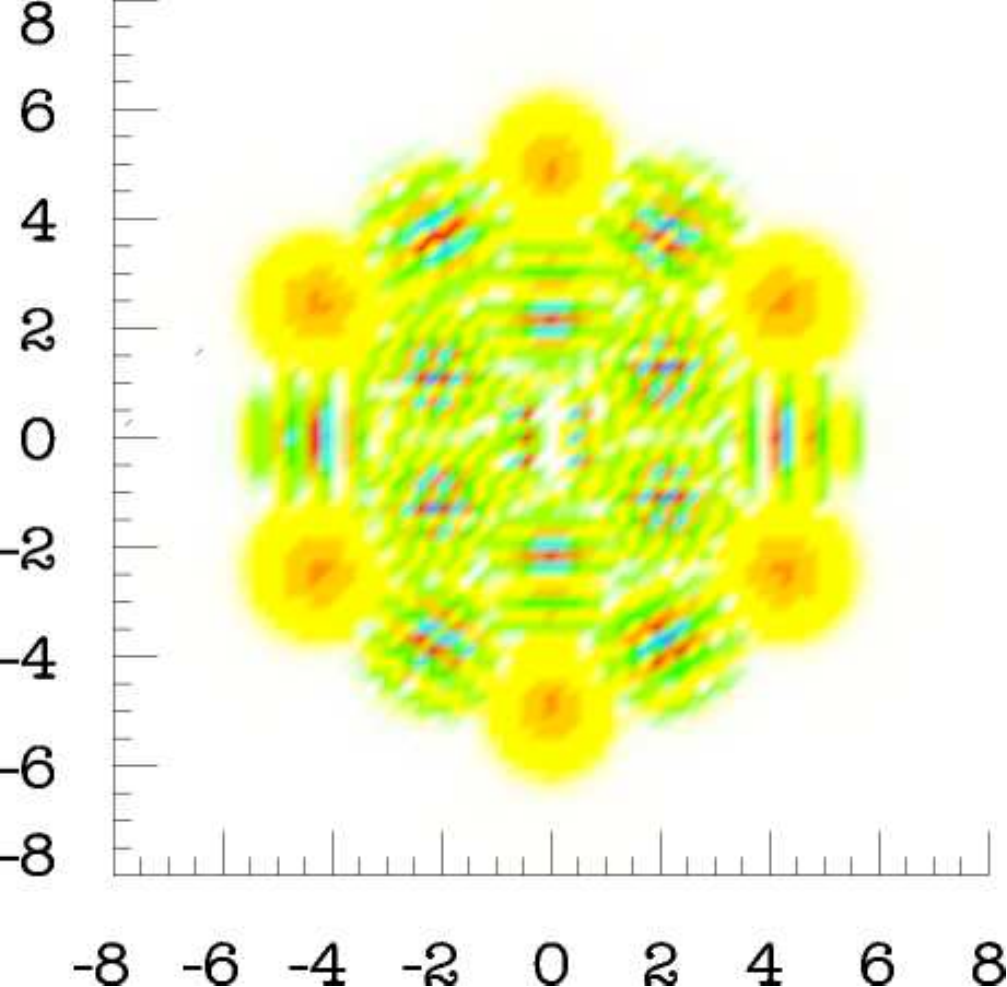}}
  \scalebox{0.5}{\includegraphics{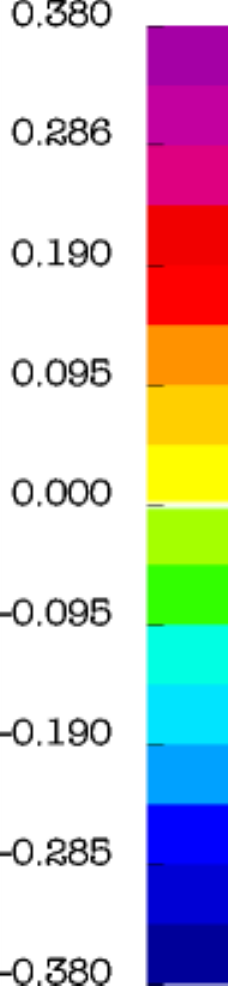}}
\end{center}
\caption{(Color online) For the evolution parameter values close to
  $\tau = 2\pi R$, where $R<1$ is a rational number, in some ares a
  destructive and in the other a constructive interference starts to
  dominate -- the left ($\tau=0.6$) and the middle figure ($\tau=1$),
  thus leading to a Wigner function of a coherent superposition state
  -- the right figure ($\tau=\pi/3$). Figures evaluated for
  $\alpha=5$.}
\label{Fig:3}
\end{figure*}

\begin{figure*}[h!]
\begin{center}
   \scalebox{0.4}{\includegraphics{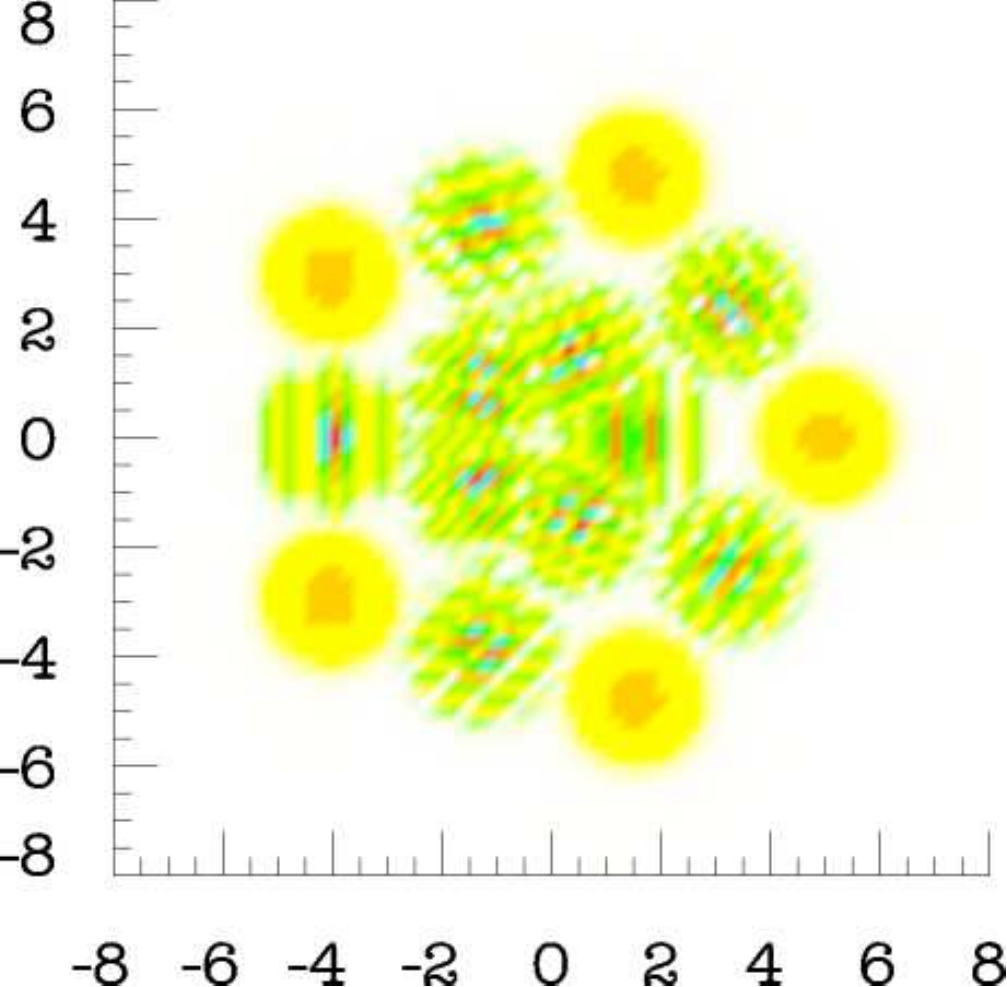}}
   \scalebox{0.4}{\includegraphics{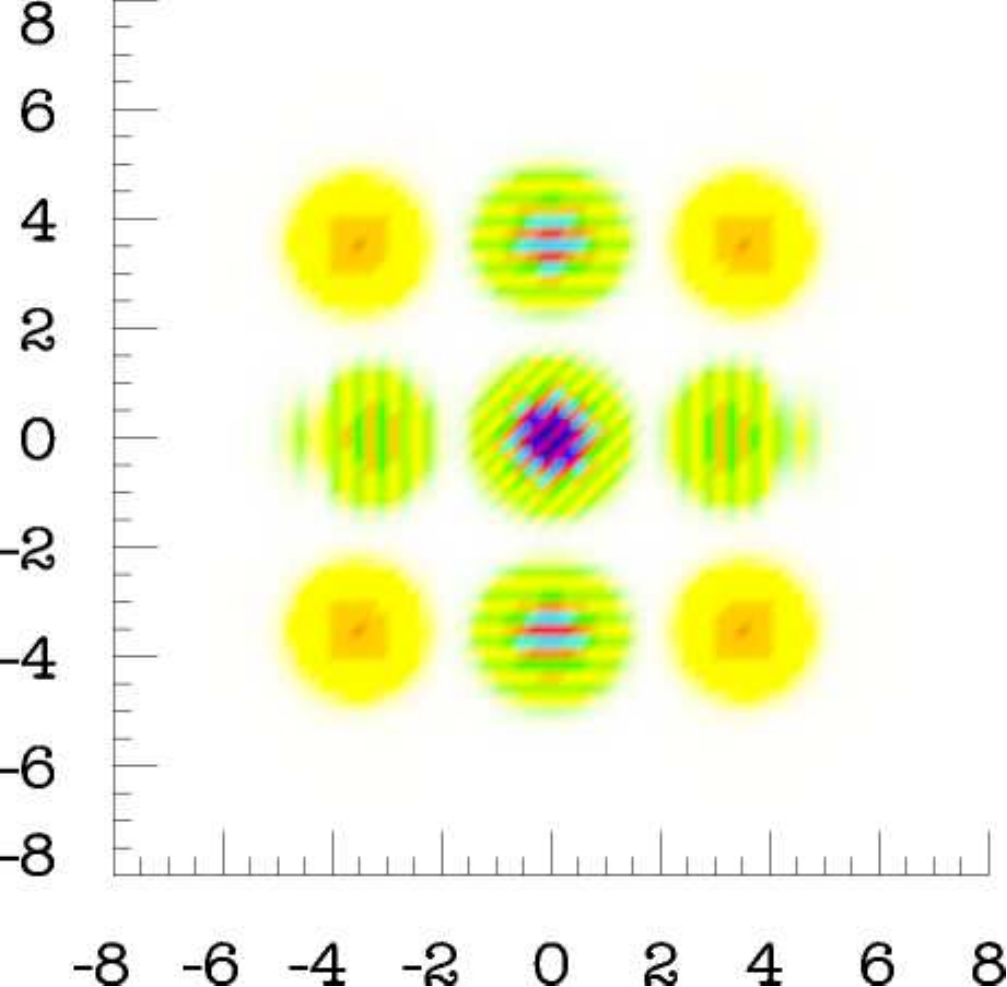}}
   \scalebox{0.4}{\includegraphics{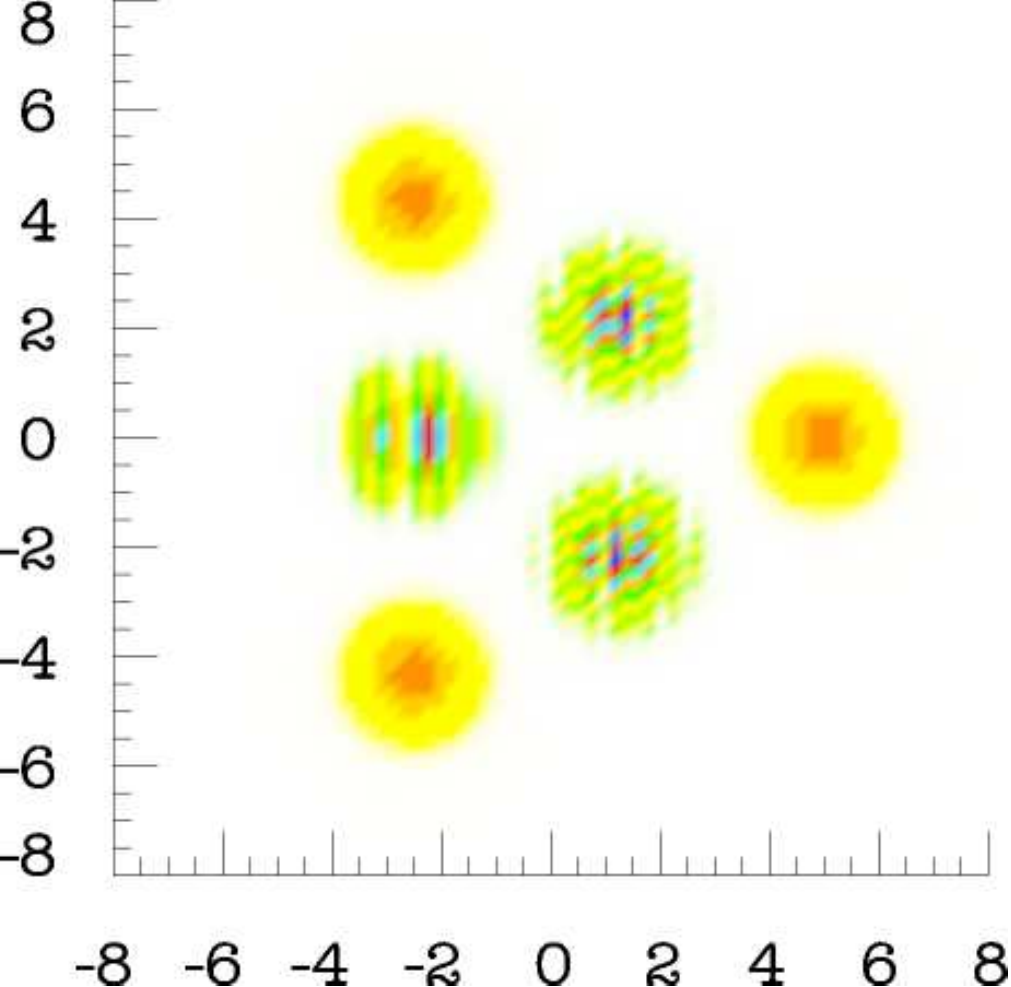}}
   \scalebox{0.4}{\includegraphics{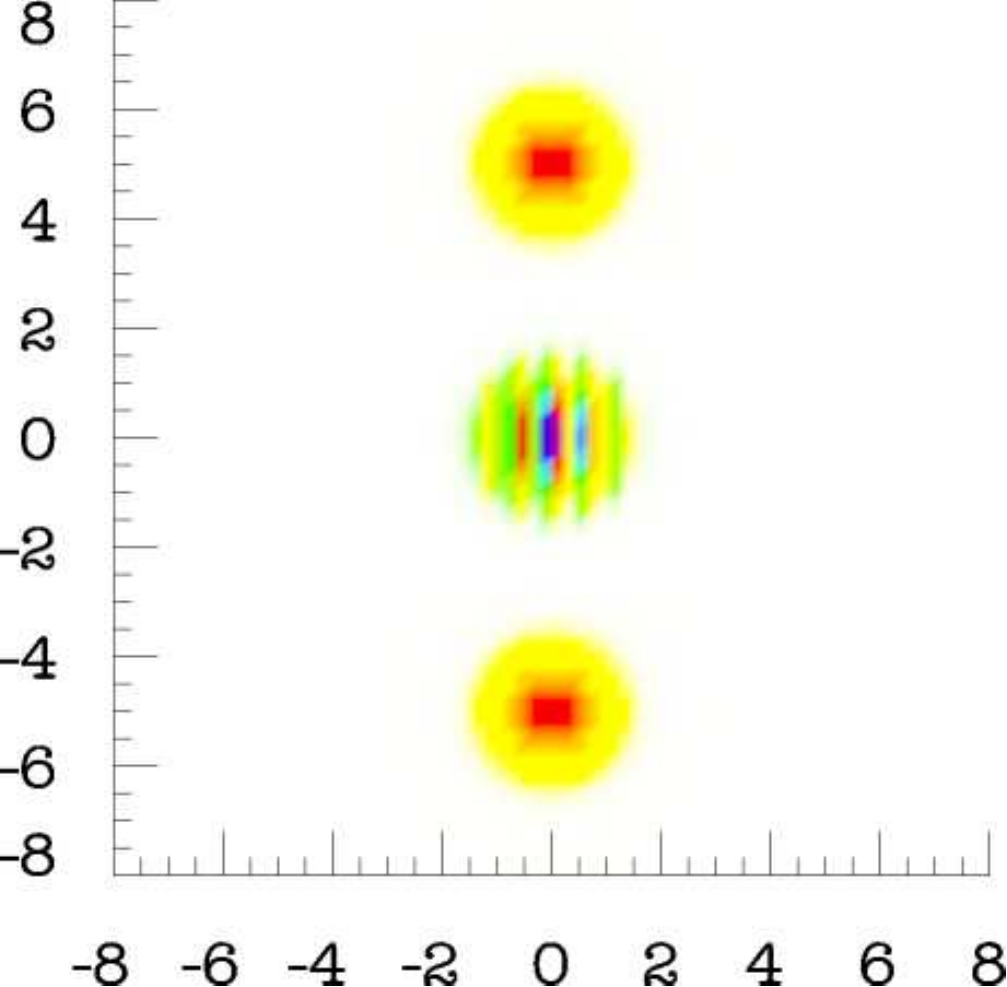}}
   \scalebox{0.4}{\includegraphics{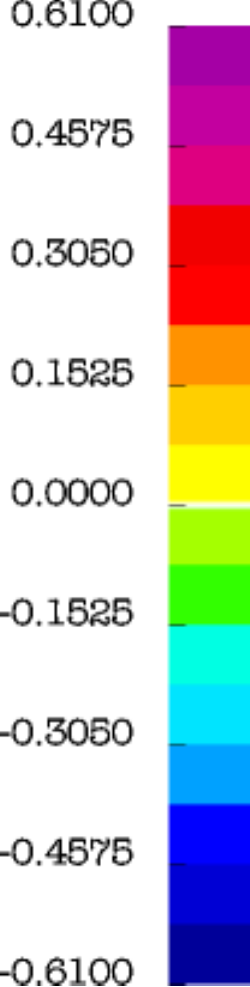}}
\end{center}
\caption{(Color online) The superposition states obtained for
  $\alpha=5$ and $\tau=2 \pi/5$ -- the left figure, $\tau=\pi/2$ (the
  compass state) -- the left middle figure, $\tau=2\pi/3$ -- the right
  middle figure and $\tau=\pi$ (the Schr\"odinger cat state) -- the
  right figure.}
\label{Fig:4}
\end{figure*}

Figures \ref{Fig:1}, \ref{Fig:2}, \ref{Fig:3}, \ref{Fig:4} present the
plots of the Wigner function for different parameters $\tau$ and
$\alpha=5$. Due to the fact that the number operator $a^{\dagger}a$ is
a constant of motion, the probability distribution during its
evolution will remain situated within a circle of radius
$|\alpha|$. Beginning from a circular shape, the Wigner function turns
into an ellipse and squeezing appears in an appropriate direction.
Then the ellipse changes into a banana shape and a ``tail'' of the
interference fringes appears where the distribution takes the negative
values.  Squeezing increases and the state becomes non-Gaussian. The
``banana part'' of the distribution gets thinner in the radial
direction and becomes more and more smeared in the azimuthal
direction.

The rotation and stretching effects which lead to squeezing are
revealed by the first term in the first line of the
equation~(\ref{eq:Fokker-Planck}). It shows the dependence of the
angular velocity on the distance $r$ from the origin of
coordinates. This corresponds to the fact that the nonlinear
refractive index $n(\gamma)$ of Kerr medium is intensity dependent
$n(\gamma) = n_0 + n_2|\gamma|^2$.  The intensity fluctuations
modulate the nonlinear refractive index and this in turn modulates the
phase of traveling light. Photons with stronger amplitude will acquire
phase faster than the photons with smaller amplitude.

The other first line terms of this equation consisting of mixed and
third order derivatives are responsible for the interference fringes
formulation.

If $\tau$ is taken as a fraction of the period of the evolution, $\tau
= 2\pi R$ where $R<1$ is a rational number, the initial coherent state
becomes a superposition of other coherent states, of the same
amplitude but different phases, see fig.~\ref{Fig:4} \cite{OSID}. This
is is also known as a fractional revival~\cite{Averbukh1989}.

The evolution governed by the self-Kerr interaction Hamiltonian is
periodic. It is easy to see this using a unitary evolution operator
$U(\tau) = \exp\{-iHt/\hbar \}$ rather than the Fokker-Planck equation
approach.  Its action on a coherent state $|\alpha\rangle$ expressed
in the Fock state basis
\begin{equation}
|\Psi(\alpha, \tau)\rangle = U(\tau) |\alpha \rangle =
e^{-\frac{|\alpha|^2}{2}}
\sum_{n=0}^{\infty}\, \frac{\alpha^n}{\sqrt{n!}}\,
e^{i\frac{\tau}{2}n(n-1)} |n\rangle
\label{eq:kerr}
\end{equation}
shows that $|\Psi(\alpha, \tau)\rangle = |\Psi(\alpha,
\tau+2\pi)\rangle$. In particular, for $\tau=2k\pi$ where $k$ is an
integer, equation~(\ref{eq:kerr}) reduces again to the coherent
state~$|\alpha\rangle$. The same periodicity holds for the Wigner
function. This condition has been used as an additional check point in
our numerical simulations.

For the wave function given by equation~(\ref{eq:kerr}) one can
determine the Wigner function as an analytical function of the
evolution parameter $\tau$ using the
definition~\cite{Scully} \begingroup
\setlength{\arraycolsep}{0cm}
\begin{equation}
W(\tau, \gamma)\!= \!\frac{2}{\pi^2}e^{2|\gamma|^2}\!\!\int \!\!d^2\beta
\langle-\beta|\rho(\alpha,\tau)|\beta\rangle
e^{-2(\beta\gamma^*-\beta^*\gamma)},
\end{equation}
\endgroup where $\rho(\alpha,\tau)$ is the density operator of the
state $|\Psi(\alpha, \tau)\rangle$.  We apply the above equation and
express it in two equivalent ways which have been used to obtain the
numerical results for the nondissipative Kerr medium
\begingroup
\setlength{\arraycolsep}{0cm}
\begin{eqnarray}
W(\tau, \gamma) &=& \frac{2}{\pi}\, e^{-2|\gamma|^2} e^{-|\alpha|^2}\,
\sum_{q=0}^{\infty} \frac{\left(2\alpha^{*}\gamma
  e^{i\frac{\tau}{2}}\right)^q}{q!}  e^{-i\frac{\tau}{2}q^2}
\nonumber\\
&\times&
\sum_{k=0}^{\infty}\frac{\left(2\alpha\gamma^*e^{-i\frac{\tau}{2}}\right)^k}
    {k!} e^{i\frac{\tau}{2}k^2} e^{-|\alpha|^2 e^{i\tau(k-q)}},
\label{summation}
\\
W(\tau, \gamma) &=& \frac{2}{\pi}\, e^{2|\gamma|^2}
e^{-|\alpha|^2}\,\sum_{n,m=0}^{\infty} \frac{1}{(-2)^{n+m}}
\frac{\alpha^n}{n!}\frac{{\alpha^*}^m}{m!}
\nonumber\\
&\times& e^{i\frac{\tau}{2}\left[n(n-1)-m(m-1)\right]}
\left(\partial_{\gamma}\right)^n \left(\partial_{\gamma^*}\right)^m
e^{-4|\gamma|^2}.  \nonumber\\
\label{derivatives}
\end{eqnarray}
\endgroup

The equations~(\ref{eq:kerr}) and~(\ref{derivatives}) explain how the
interference fringes arise from the interference of the Fock number
states $|n\rangle$.  By virtue of equation~(\ref{derivatives}), the
Wigner function can be viewed as a sum of the two components: the
first one, independent of $\tau$, being a superposition of the Fock
states' Wigner functions $W_{|n\rangle \langle n|}(\gamma)$ and the
second, being an evolution dependent component $f(\tau, \gamma)$,
responsible for the evolution of the fringes
\begin{equation}
W(\tau,\gamma) = \frac{e^{-|\alpha|^2}}{\pi} \sum_{n=0}^{\infty}
\frac{|\alpha|^{2n}}{n!}\; W_{|n\rangle \langle n|}(\gamma) + f(\tau,
\gamma).
\end{equation}

Although the equation (\ref{summation}) and (\ref{derivatives}) are
special solutions of the Fokker-Planck equation, we obtained them
separately to proceed independent computation.

The effects of squeezing and error contour rotation in the phase space
can also be observed by studying the easy-to-compute Q-function
evolution $Q(\tau, \gamma) = 1/\pi \langle\gamma |\rho(\alpha, \tau)
|\gamma\rangle$
\begingroup \setlength{\arraycolsep}{0cm}
\begin{equation}
Q(\tau, \gamma) = \frac{1}{\pi} e^{-|\alpha|^2-|\gamma|^2}
\left|\sum_{n} \frac{(\alpha \gamma^*)^n}{n!} \,
e^{i\frac{\tau}{2}n(n-1)}\right|^2.
\end{equation}
\endgroup The contour plot of the Q-function is plotted in
\cite{Tanas}. Solutions of the Fokker-Planck equation for the
Q-function in a dissipative and noisy Kerr medium have been widely
studied \cite{Tanas, Milburn1986, Milburn1989, Perinova1990}.
Negativities achieved in the Wigner function correspond to zeros of
the Q-function~\cite{Korsch1997}.

All the figures \ref{Fig:1}--\ref{Fig:4} have been obtained computing
the equations (\ref{eq:Fokker-Planck}), (\ref{summation}) and
(\ref{derivatives}) independently.

\section{Dissipation effects}\label{ts}

In the presence of damping the Wigner function will not remain
situated within at the circle of the radius $|\alpha|$ any more.  The
second line in the Fokker-Planck equation, proportional to the first
radial derivative, makes the Wigner function moving towards the origin
of the coordinates. This effect corresponds to decreasing number of
quanta in the state during the evolution.  At the origin the state
becomes vacuum.

The third line describes the effects of dispersion.

It is interesting to note an interplay between the nonlinear evolution
and damping terms in the Fokker-Planck equation.  If damping is
negligible, the numerical simulations obtained for small $|\alpha|
\simeq 1$ and large $|\alpha| \gg 1$ initial amplitudes are very
similar. At the beginning of evolution the squeezing effect dominates
and then the interference appears.  It is not the case if the losses
are included in the simulations.  If the amplitude is too small the
decoherence washes out all the quantum effects and the Gaussian Wigner
function located at the coordinates' origin, genuine to vacuum state,
is obtained very quickly.

We proceed the numerical computation for experimentally reasonable
values genuine to a nanomechanical resonator.  The nonlinear constant
$\kappa\simeq 2\cdot 10^{-4}\mathrm{s^{-1}}$ and the damping constant
$\Gamma \simeq 10^3 \, \mathrm{s^{-1}}$ ($\xi=10^5$).  The thermal
noise coefficient in the room temperature is equal to $N \simeq
1.9\cdot 10^{-14}$.

The nonlinear evolution term $r^2$ for $\tau \simeq 0$ becomes
meaningful if $r \simeq |\alpha|$.  If $|\alpha| \simeq 1$ the damping
term $\frac{\xi r}{2} \simeq 0.5 \cdot 10^5$ dominates over the
nonlinear $r^2 \simeq 1$ and interference terms. The coherent state
will be almost immediately radially displaced towards the origin of
coordinates.  For $|\alpha| \simeq 10^5$ all the effects are in
balance: $r^2 \simeq 10^{10}$ and $\frac{\xi r}{2} \simeq 0.5 \cdot
10^{10}$.

Computing the Fokker-Planck equation for such large values of
coefficients ($\alpha \simeq 10^5$, $\xi \simeq 10^5$, $N=1.9\cdot
10^{-14}$) would require a very dense and large grid, thus a lot of
computer memory. Therefore, to visualize an influence of these two
effects on the Wigner function, we rescaled the parameters.

As we pointed out above, in order to see the nonlinear effects during
the Wigner function evolution in presence of dissipation, the input of
nonlinear term has to be of the same order as the dissipation term is.
Therefore, we keep the ratio between them constant and equal to the
ratio evaluated for the non-rescaled case: $\xi\simeq \alpha$ and $N =
1.9\cdot 10^{-19} \xi$. For $\alpha =2$ they are equal to $\xi \simeq
2$ and $N = 3.8 \cdot 10^{-19}$.  Although such a small value of the
thermal coefficient implies a negligible effect on the evolution, we
included it in the simulation.  Such simulations are accessible using
a standard PC computer with 4 GB of memory.

\begin{figure*}[h!]
\begin{center}
   \scalebox{0.4}{\includegraphics{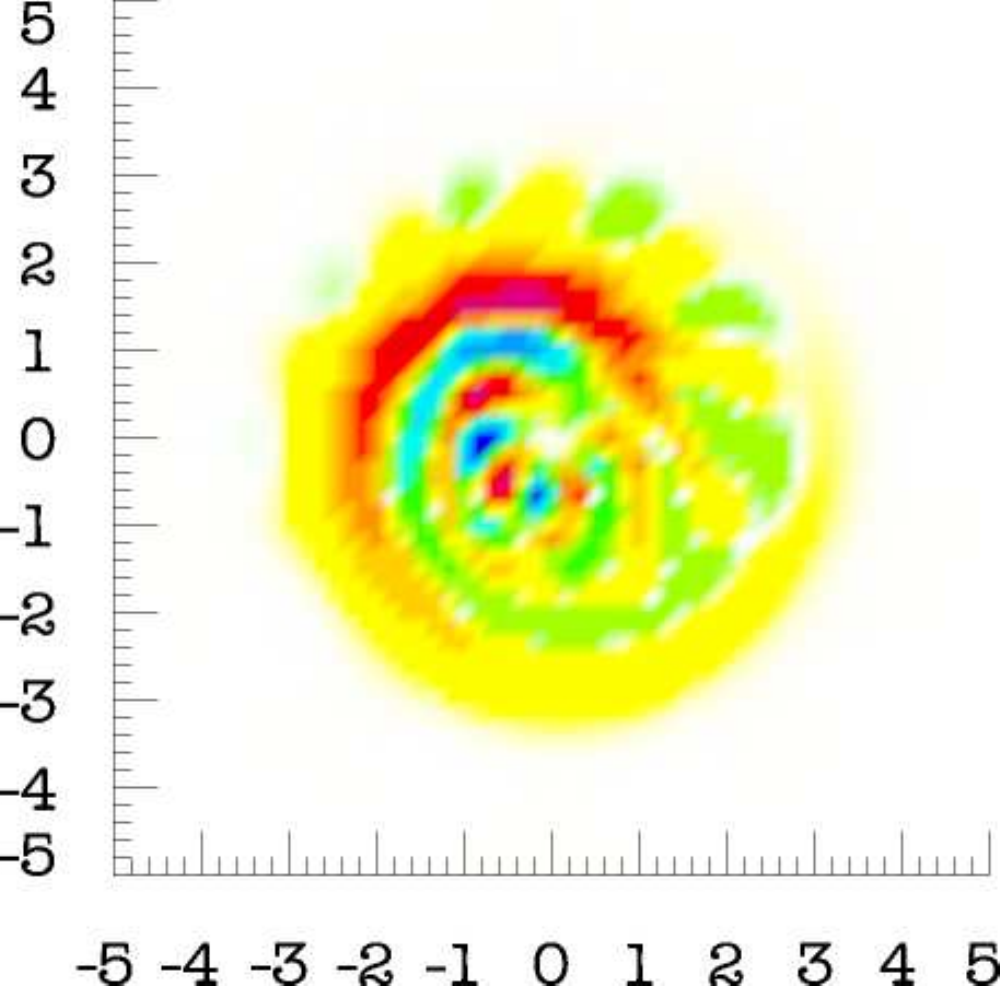}}
   \scalebox{0.4}{\includegraphics{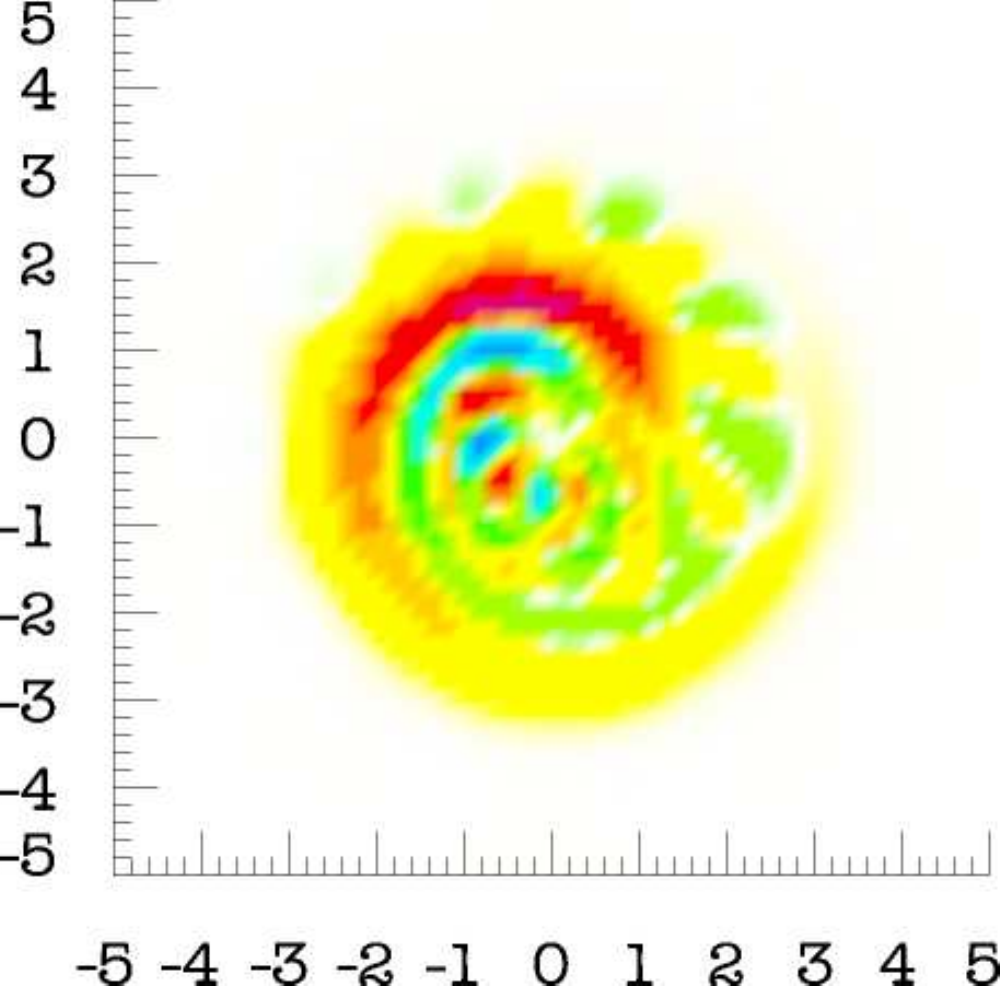}}
   \scalebox{0.4}{\includegraphics{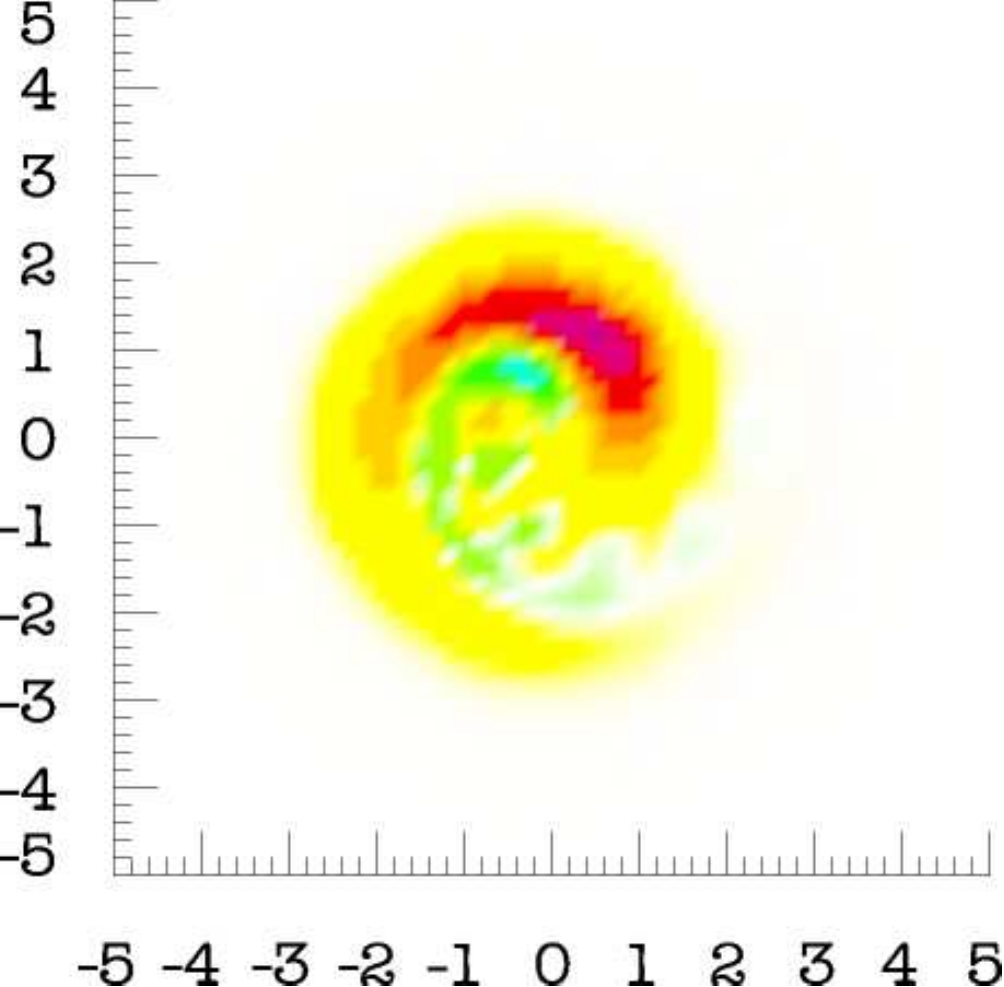}}
   \scalebox{0.4}{\includegraphics{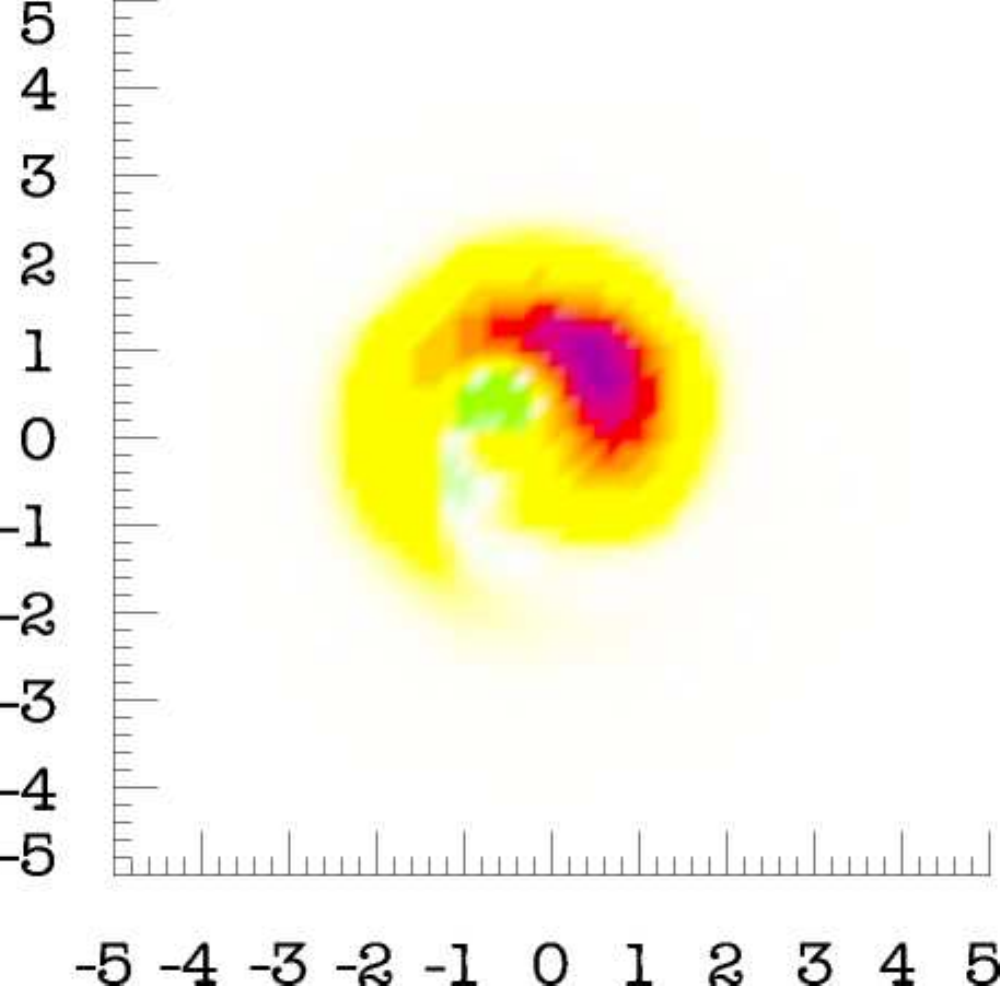}}
   \scalebox{0.4}{\includegraphics{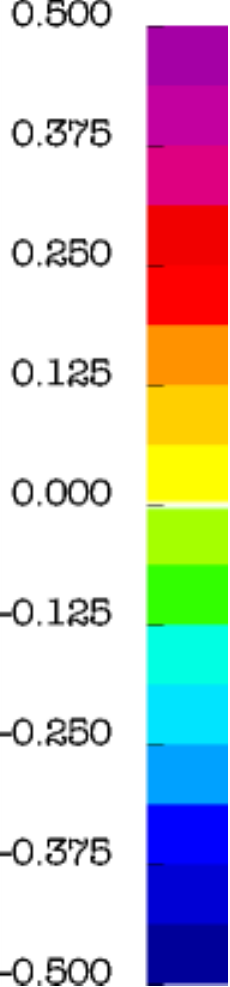}}
\end{center}
\caption{(Color online) The effect of the decoherence on the Wigner function
  evaluated for $\alpha=2$, $\tau=0.2\pi$, the thermal noise $N=3.8
  \cdot 10^{-19}$ and the different values of the damping constant:
  $\xi=0$ -- the left figure, $\xi=0.1$ -- the left middle figure,
  $\xi=1$ -- the right middle figure, $\xi=2$ -- the right
  figure. Note that the thermal noise gives no input if $\xi=0$.}
\label{Fig:5}
\end{figure*}

\begin{figure*}[h!]
\begin{center}
   \scalebox{0.4}{\includegraphics{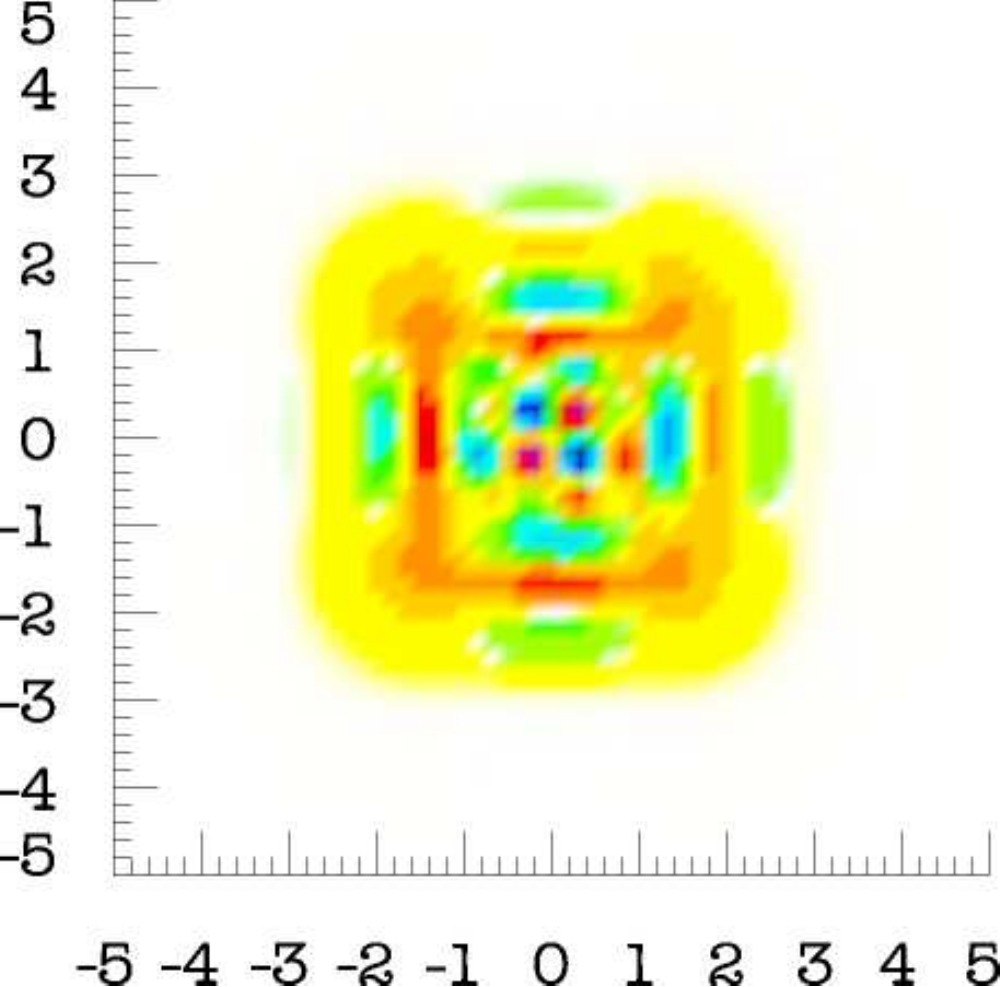}}
   \scalebox{0.4}{\includegraphics{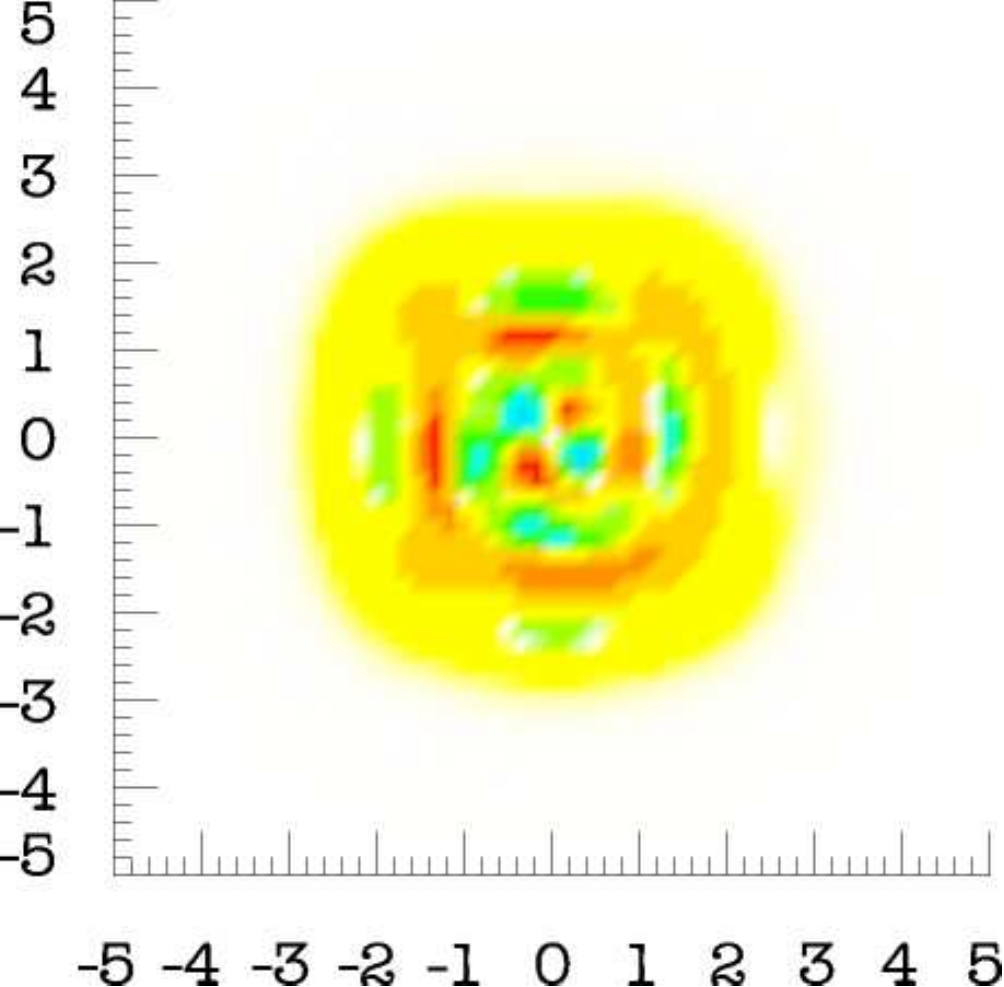}}
   \scalebox{0.4}{\includegraphics{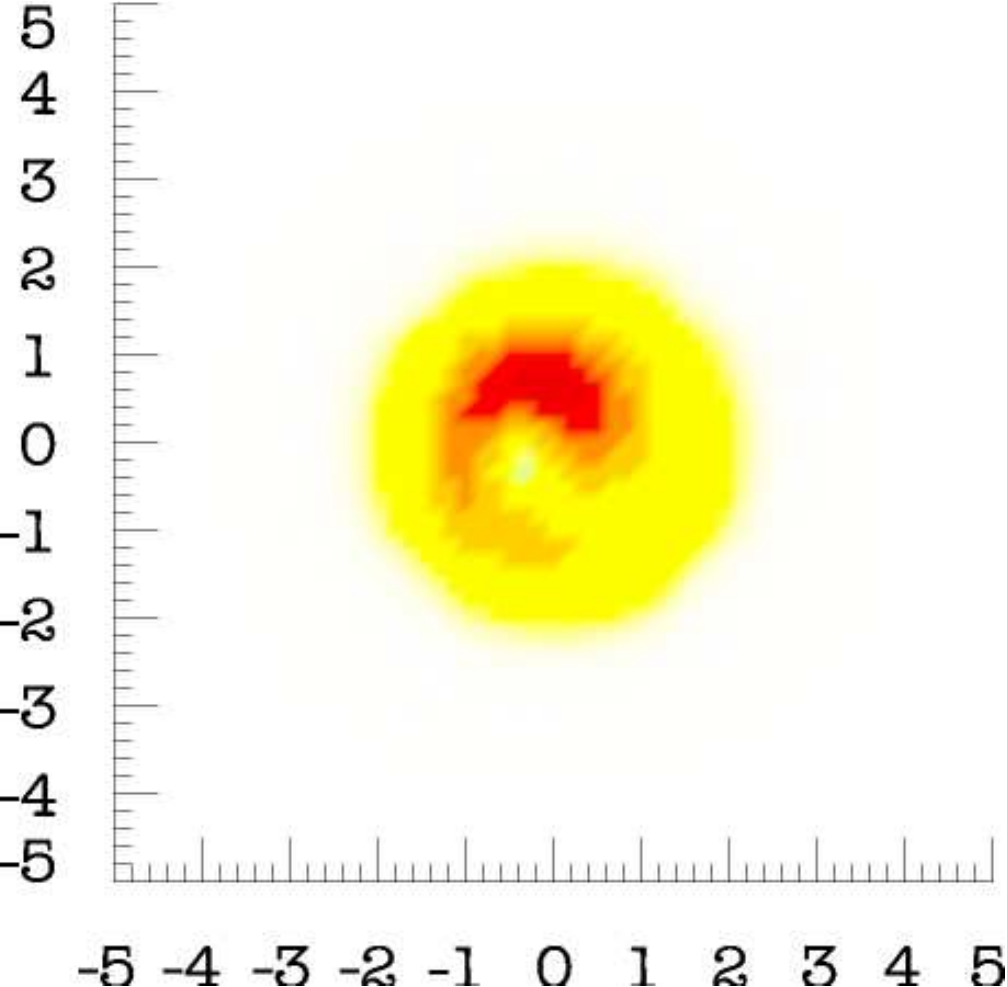}}
   \scalebox{0.4}{\includegraphics{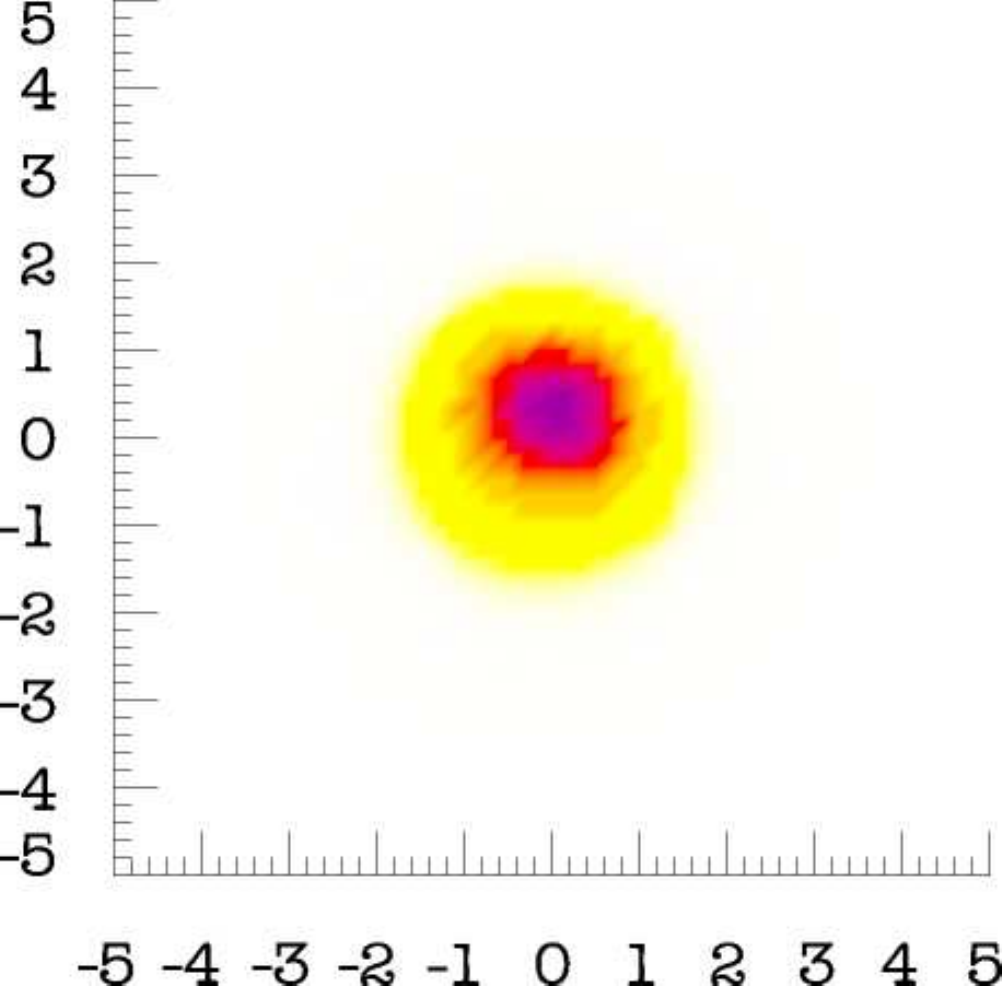}}
   \scalebox{0.4}{\includegraphics{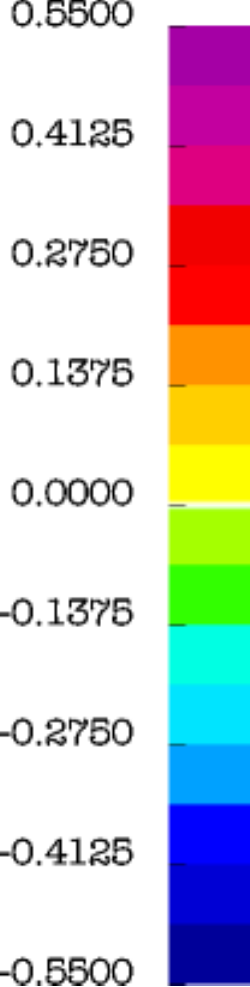}}
\end{center}
\caption{(Color online) The coherent superposition states are extremely fragile to
  the decoherence process. The Wigner function of four coherent state
  superposition ($\tau=\pi/2$) of the amplitude $\alpha=2$
  evaluated including the thermal noise $N=3.8 \cdot 10^{-19}$ and the
  damping constant equal to $\xi=0$ -- the left figure, $\xi=0.1$ --
  the left middle figure, $\xi=1$ -- the right middle figure, $\xi=2$
  -- the right figure. }
\label{Fig:6}
\end{figure*}

\begin{figure*}[h!]
\begin{center}
   \scalebox{0.4}{\includegraphics{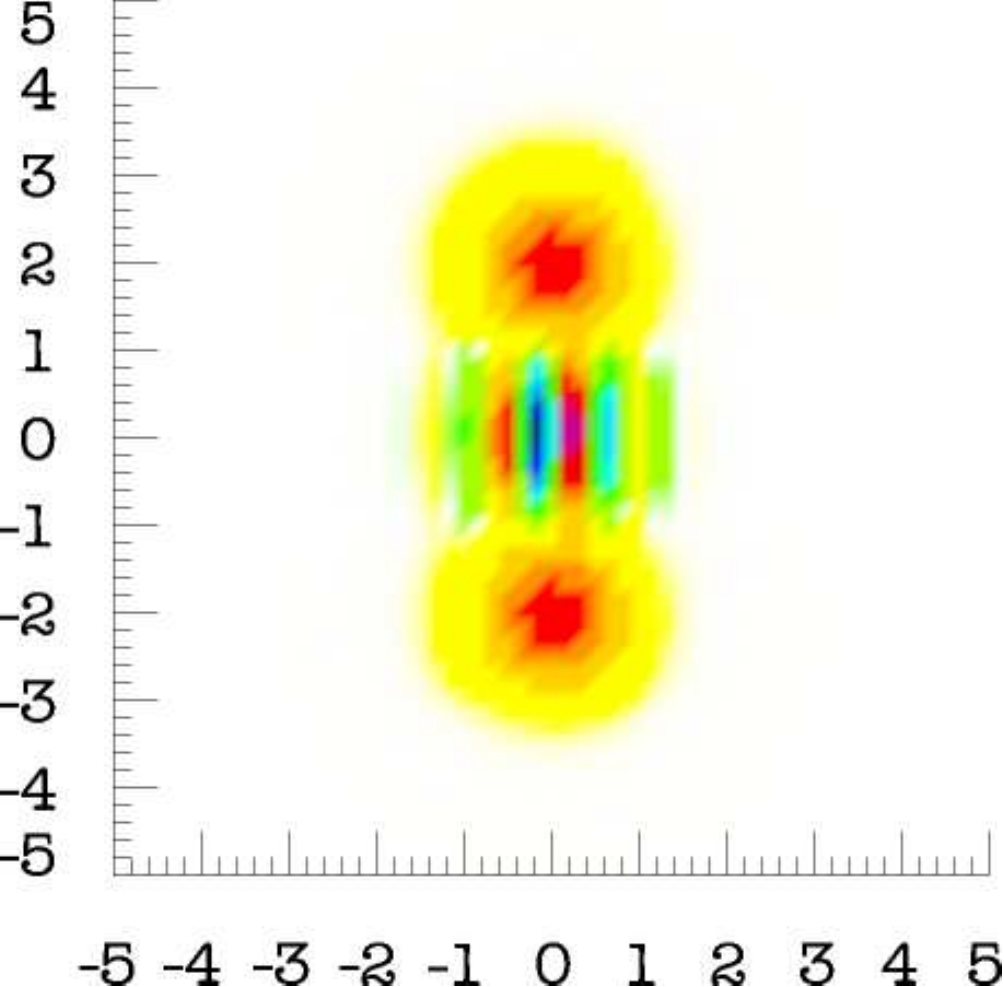}}
   \scalebox{0.4}{\includegraphics{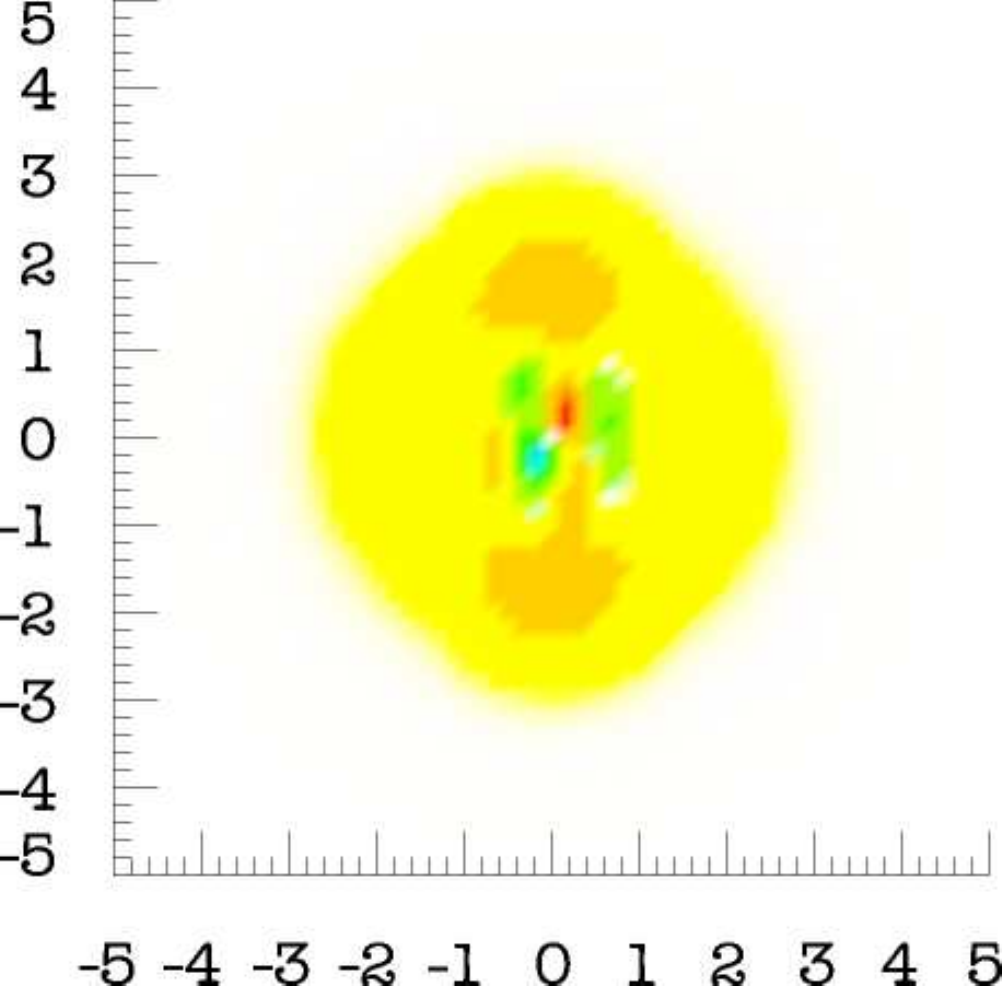}}
   \scalebox{0.4}{\includegraphics{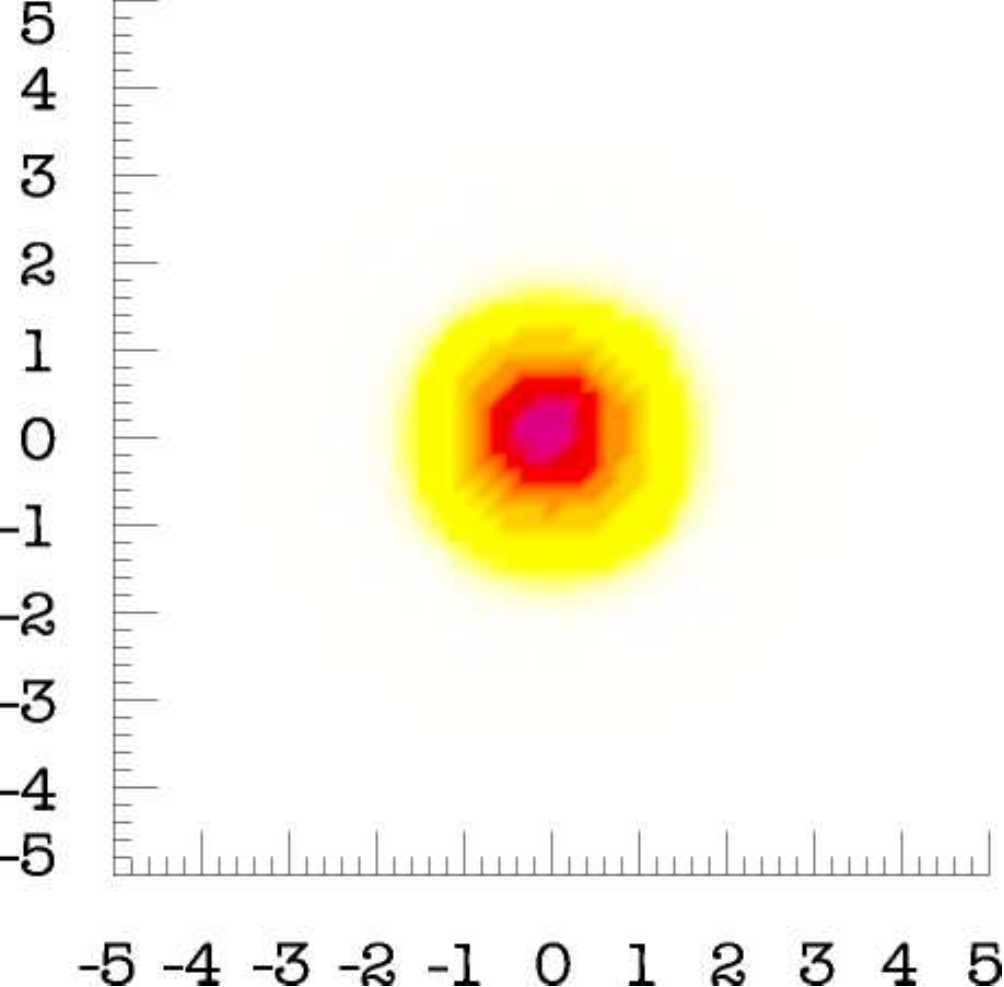}}
   \scalebox{0.4}{\includegraphics{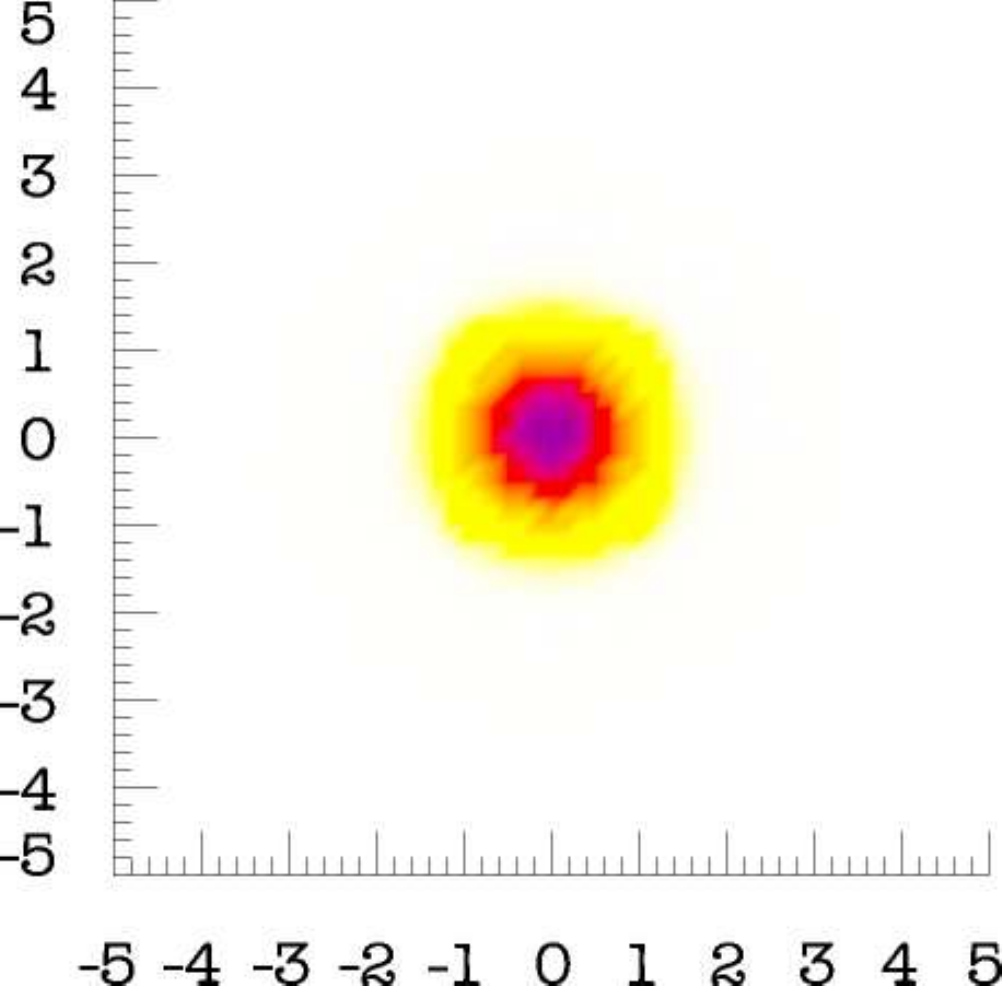}}
   \scalebox{0.4}{\includegraphics{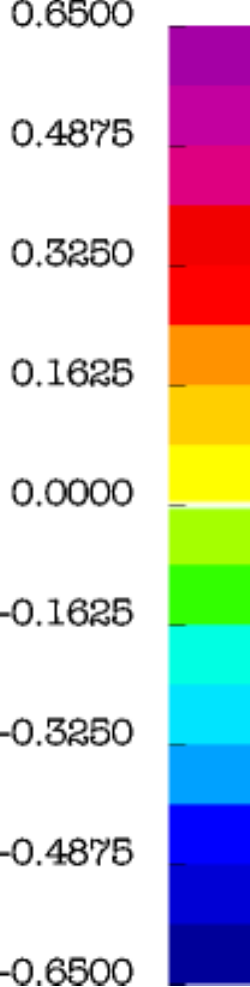}}
\end{center}
\caption{(Color online) The decoherence looks similarly for the two coherent state
  superposition Wigner function ($\tau=\pi$), also of amplitude
  $\alpha=2$, the same thermal noise $N=3.8 \cdot 10^{-19}$ and the
  damping constants: $\xi=0$ -- the left figure, $\xi=0.1$ -- the left
  middle figure, $\xi=1$ -- the right middle figure, $\xi=2$ -- the
  right figure.}
\label{Fig:7}
\end{figure*}

\begin{figure*}[h!]
\begin{center}
   \scalebox{0.4}{\includegraphics{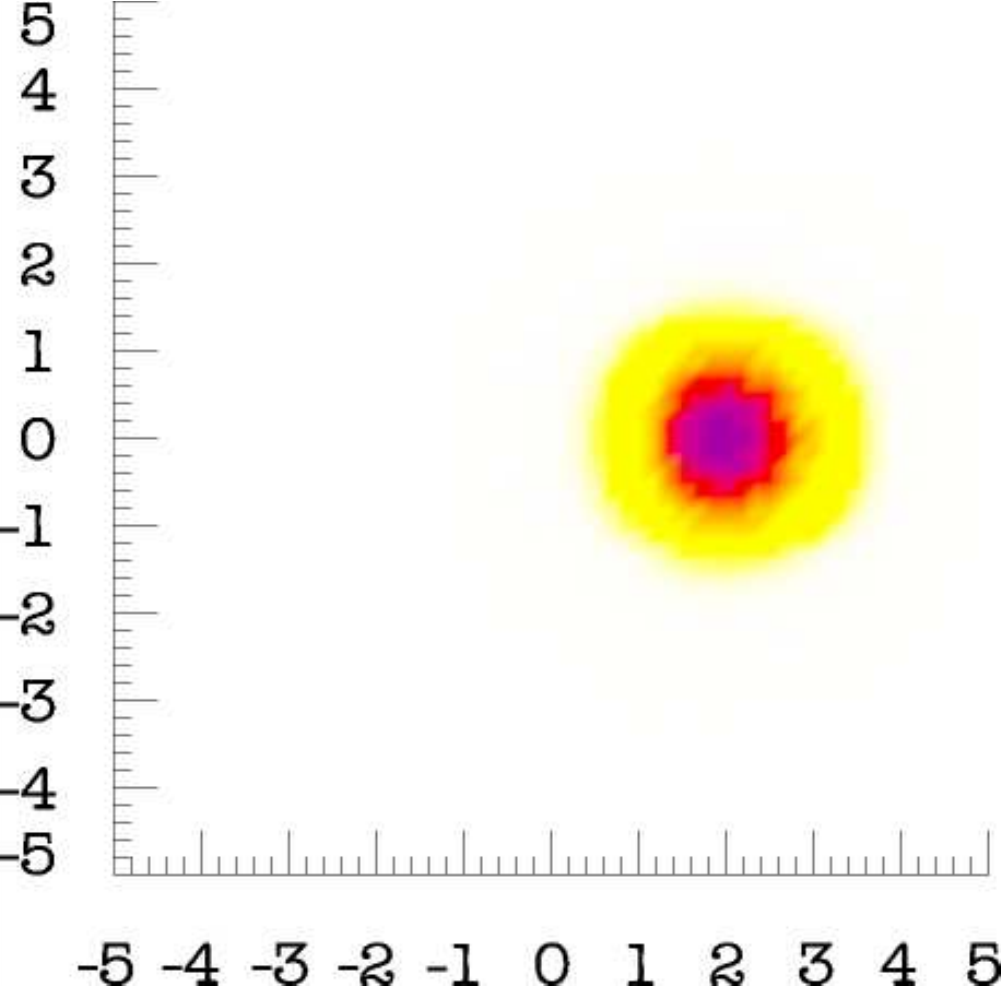}}
   \scalebox{0.4}{\includegraphics{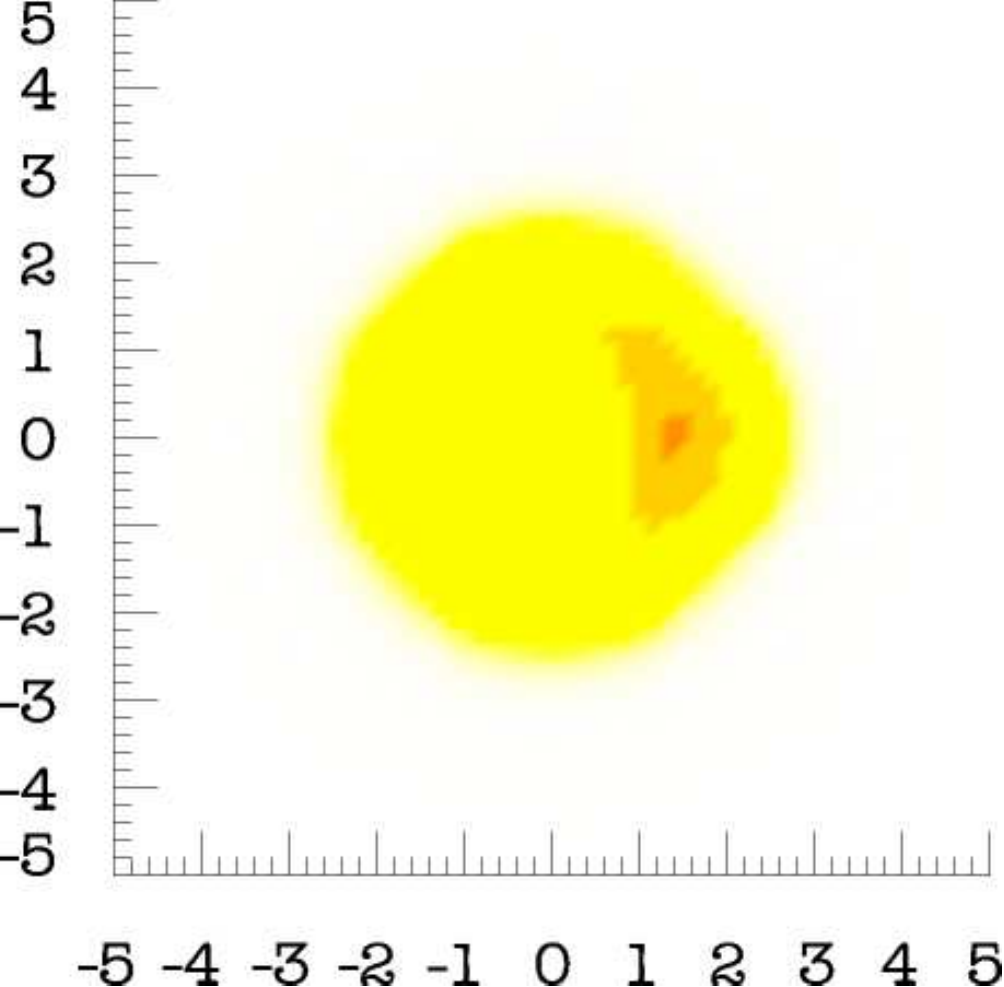}}
   \scalebox{0.4}{\includegraphics{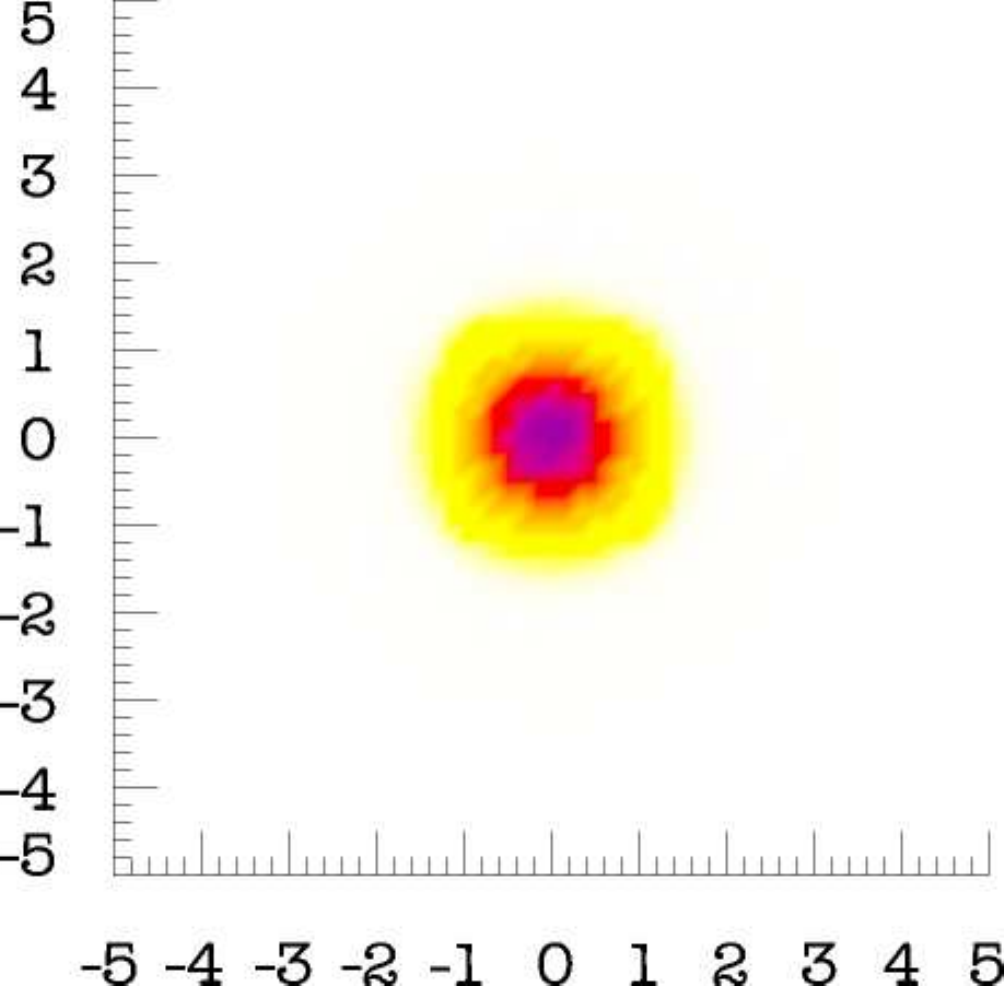}}
   \scalebox{0.4}{\includegraphics{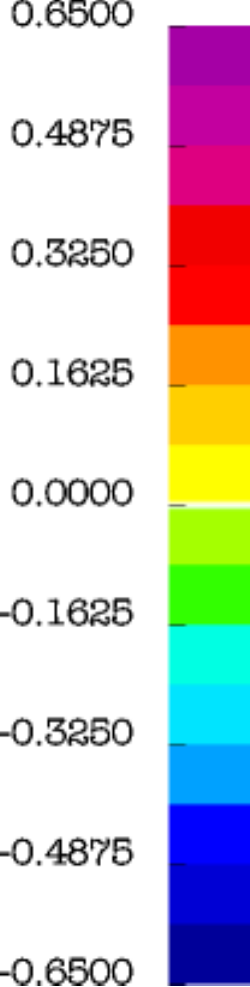}}
\end{center}
\caption{(Color online) The Wigner function evolution is a noisy $\chi^{(3)}$ medium
  is not periodic any more. The distribution evaluated for $\alpha=2$,
  $\tau=2\pi$, the thermal noise $N=3.8 \cdot 10^{-19}$ and the damping
  constant equal to $\xi=0$ -- the left figure, $\xi=0.1$ -- the
  middle figure, $\xi=1$ -- the right figure. The state pursues
  vacuum.}
\label{Fig:8}
\end{figure*}

Figures \ref{Fig:5}, \ref{Fig:6}, \ref{Fig:7}, \ref{Fig:8} present the
numerical simulations of the Fokker-Planck equation obtained for
$\alpha =2$, $N = 3.8 \cdot 10^{-19}$ and the following values of the
damping constant: $\xi=0$, $\xi=0.1$, $\xi=1$, $\xi=2$.

Due to the fact that damping washes out the interference effects, any
nonzero damping constant prevents both: coherent superposition state
formulation and periodicity of the evolution. As it is well
known~\cite{Zurek} these states are extremely fragile to the
decoherence.  The areas where the Wigner function takes the negative
values disappear, see fig.~\ref{Fig:5}. For an ideal $\chi^{(3)}$
medium the negativities show up for $0.08 +2k\pi < \tau < 6.20+ 2k\pi$
where $k$ is an integer. If $\xi=0.1$ the negativities are present for
$0.08 < \tau < 5.80$. However, they do not appear in the second round
($0.08 + 2\pi< \tau < 5.80 + 2\pi$). If $\xi=1$ the negative values
are present only for $0.08 < \tau < 1.52$ and for $0.08 < \tau < 0.89$
if $\xi=2$.

The Wigner function obtained for the coherent superposition states
evolution times $\tau = \frac{\pi}{2}$, $\tau = \pi$ and $\tau =2\pi$
is depicted on figures \ref{Fig:6}, \ref{Fig:7}, \ref{Fig:8}. For $\xi
= 0.1$ the structure of superposition coherent states is still well
preserved. However, some additional circular trails are present. These
are specially visible for $\tau=\pi$ and $\tau=2\pi$. For $\xi=2$ and
$\tau = \frac{\pi}{2}$ the state is already a vacuum.  The vacuum
state is achieved for $\tau = \pi$ ($\tau = 10\pi$) if $\xi=1$
($\xi=0.1$).

\section{Sub-Planck structure in phase-space} \label{sub}

In the framework of quantum mechanics, it has been recognized that
small structures on the sub-Planck scale do show up in quantum linear
superpositions \cite{ZurekN, ZurekPRA}. Such sub-Planck structures can
be shown, if the Wigner function is plotted in a phase-space region
below the Heisenberg relation.  That means in the phase space volume
less than $1$, since $\Delta X_1 \Delta X_2 \ge 1$ for the amplitude
$X_1$ and phase $X_2$ quadratures.  The quadratures are given by the
annihilation $a = \frac{1}{2} (X_1 + i X_2)$ and creation $a^{\dagger}
= \frac{1}{2} (X_1 - i X_2)$ operators and their mean values
correspond to phase-space coordinates in the following way $\langle
X_1\rangle = \mathrm{Re} \, \gamma$, $\langle X_2\rangle = \mathrm{Im}
\, \gamma$.

\begin{figure*}[h!]
\begin{center}\hbox to \hsize{\hss
   \scalebox{0.4}{\includegraphics{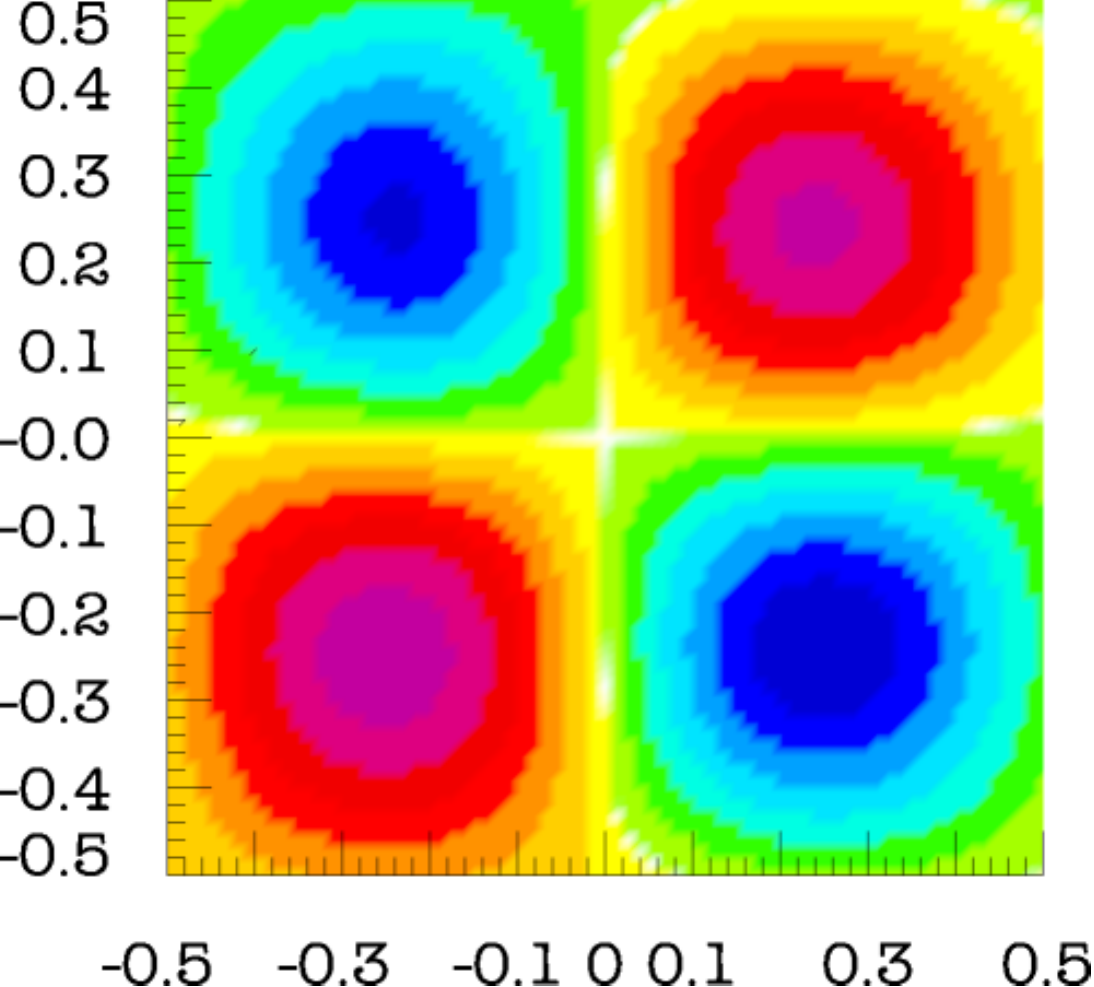}}
   \scalebox{0.4}{\includegraphics{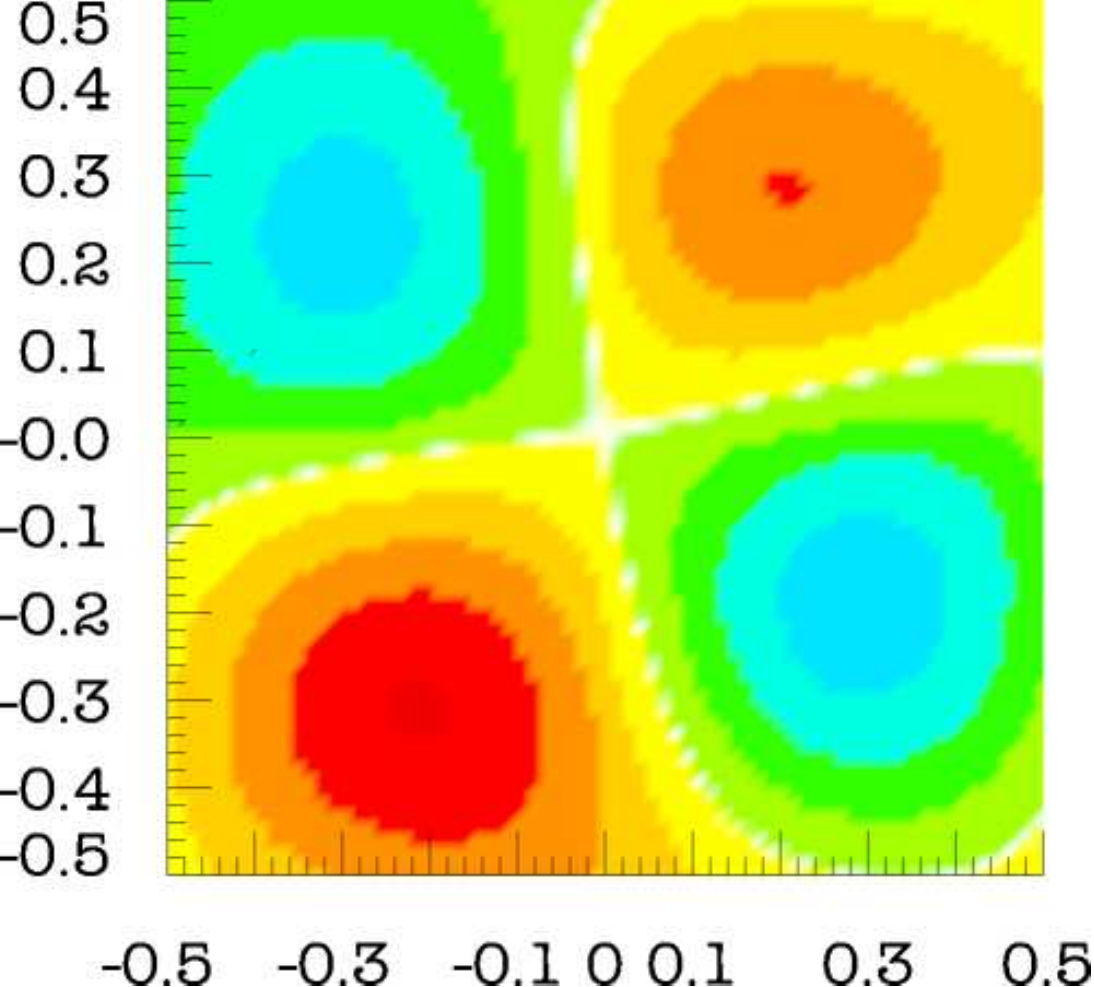}}
   \scalebox{0.4}{\includegraphics{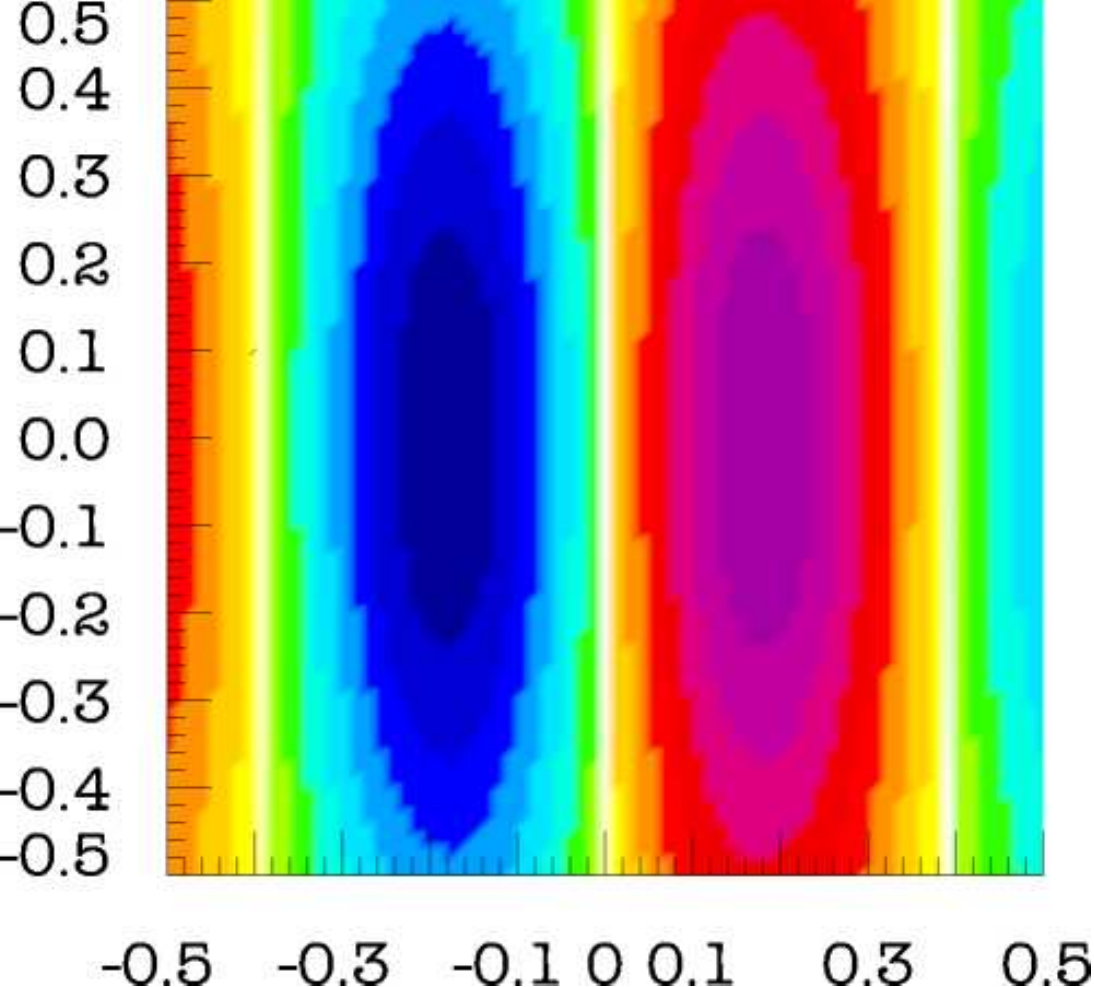}}
   \scalebox{0.4}{\includegraphics{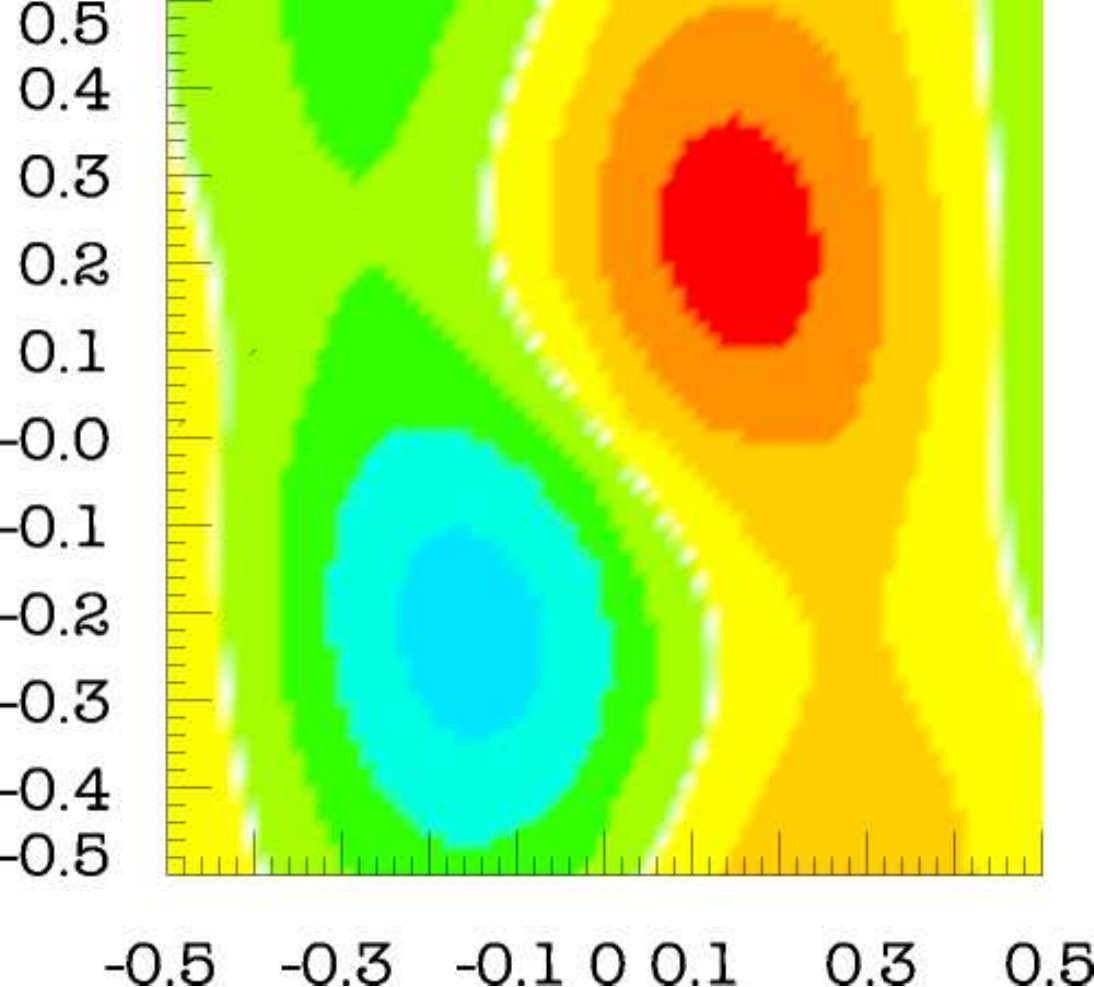}}
   \scalebox{0.4}{\includegraphics{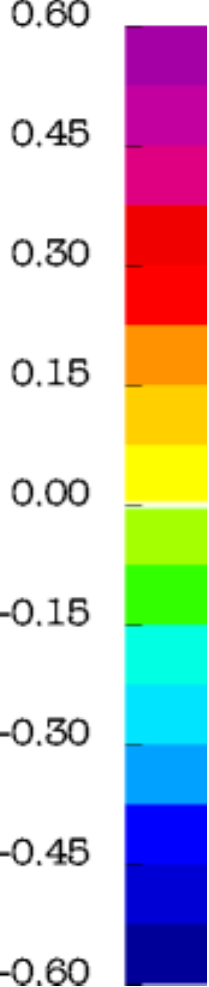}}\hss}
\end{center}
\caption{(Color online) A sub-Planck structure of the Wigner function for $\tau =
  \pi/2$ and $\xi = 0$ -- the left figure, $\tau = \pi /2$ and $\xi =
  0.1$ -- the left middle figure, $\tau = \pi$ and $\xi = 0$ -- the
  right middle figure, $\tau = \pi$ and $\xi = 0.1$ -- the right
  figure.}
\label{Fig:9}
\end{figure*}

Figure \ref{Fig:9} compares behavior of the sub-Planck structure of
the Wigner function obtained for the compass state ($\tau = \pi/2$)
and the Schr\"odinger cat state ($\tau = \pi$) for $\alpha=2$ in the
ideal $\chi^{(3)}$ medium and in presence of damping ($\xi=0.1$). The
regions of the opposite sign, in the form of ``dots,'' are clearly
visible. They become flattened and smeared by the damping effects.

\begin{figure*}[h!]
\begin{center}\hbox to \hsize{\hss
   \scalebox{0.4}{\includegraphics{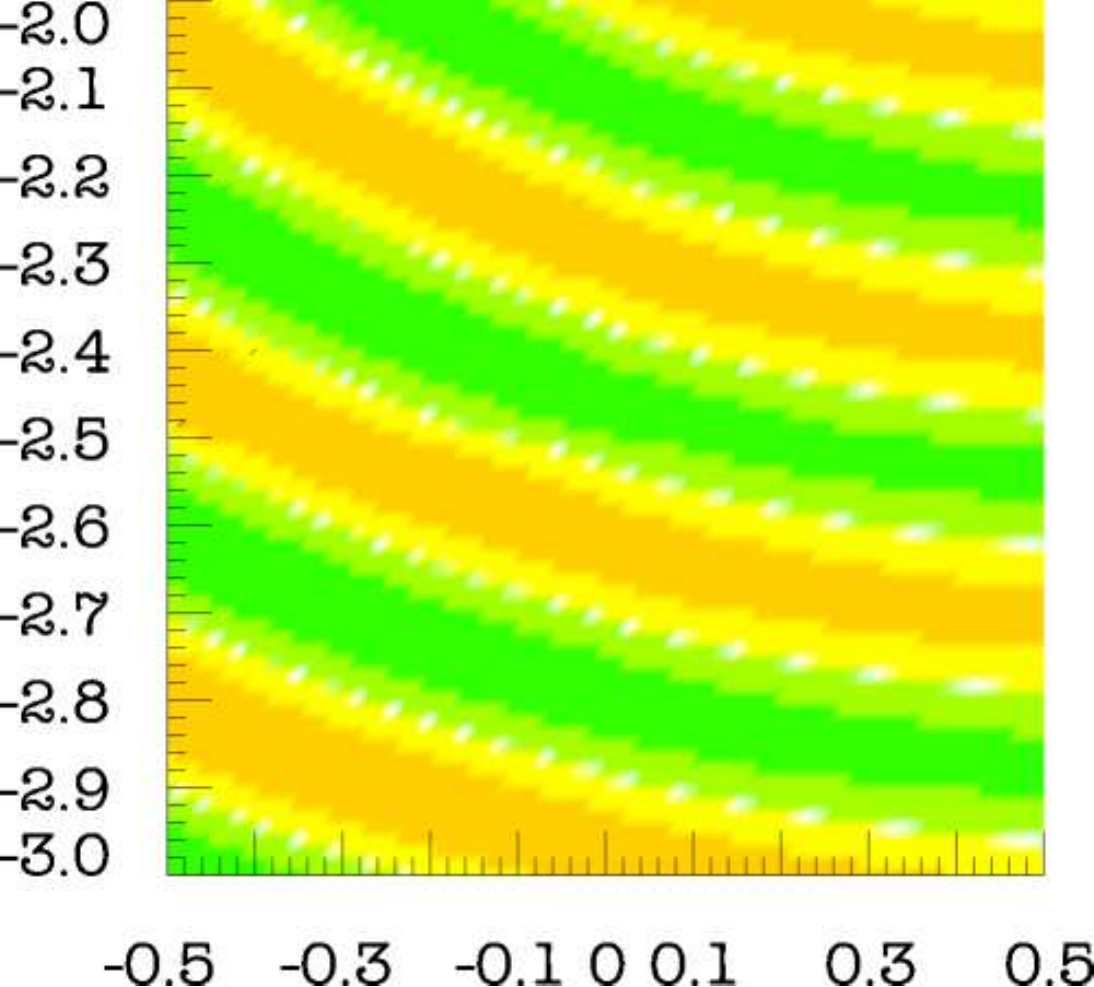}}
   \scalebox{0.4}{\includegraphics{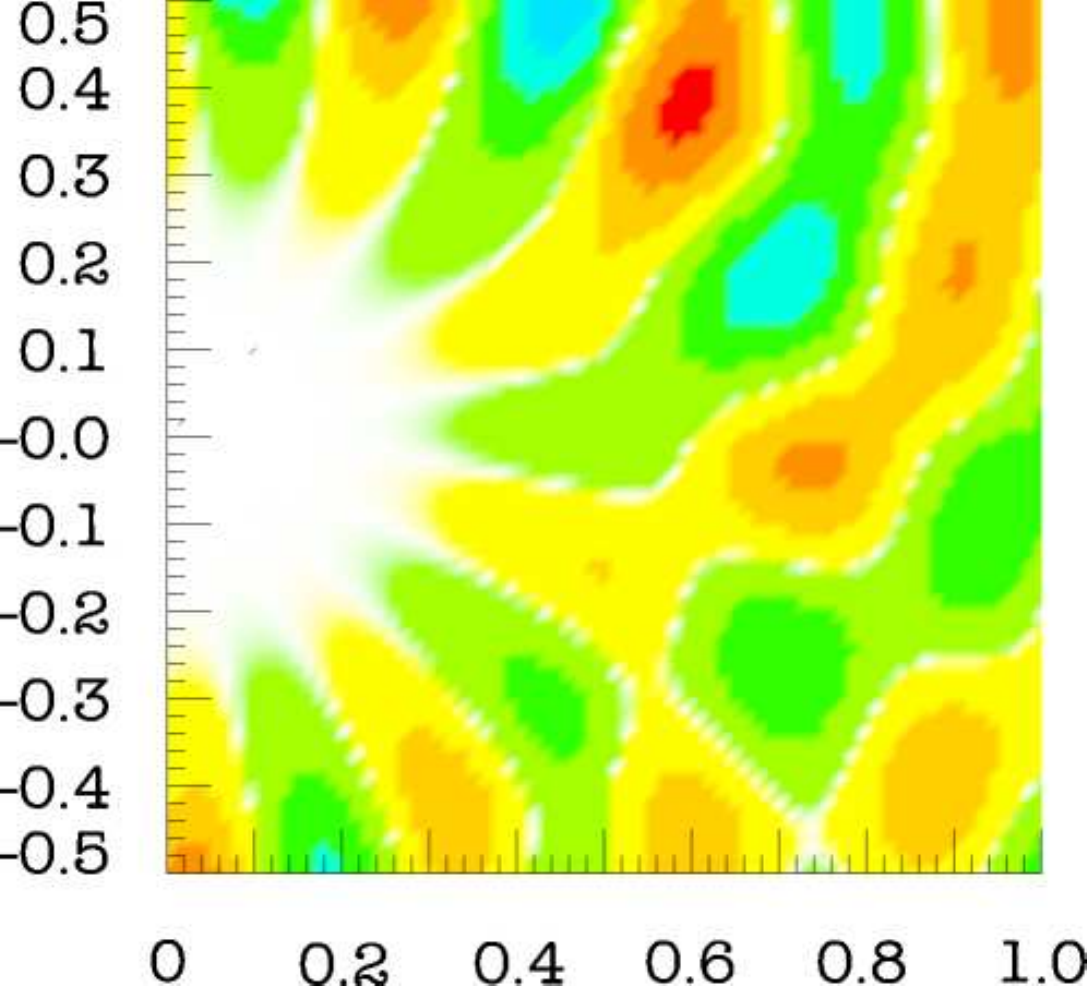}}
   \scalebox{0.4}{\includegraphics{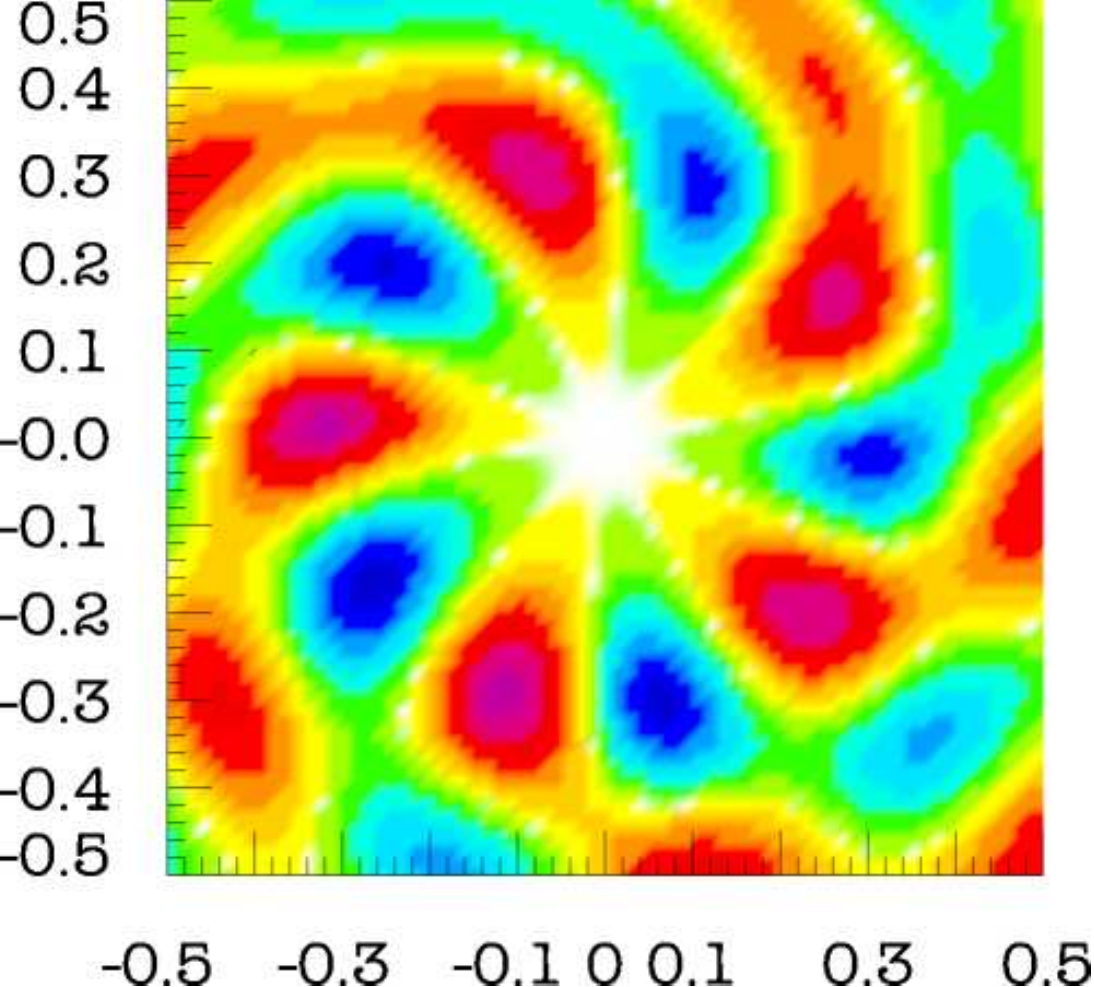}}
   \scalebox{0.4}{\includegraphics{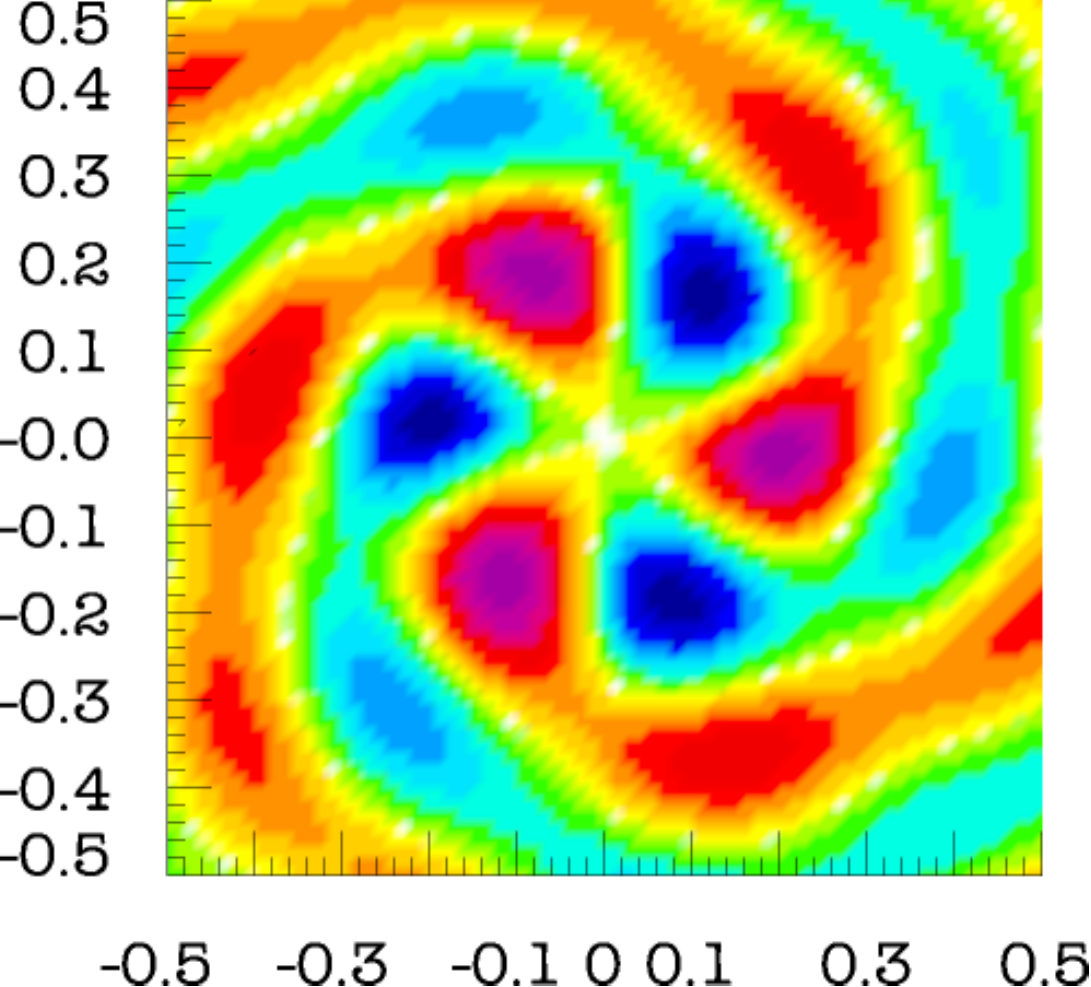}}
   \scalebox{0.4}{\includegraphics{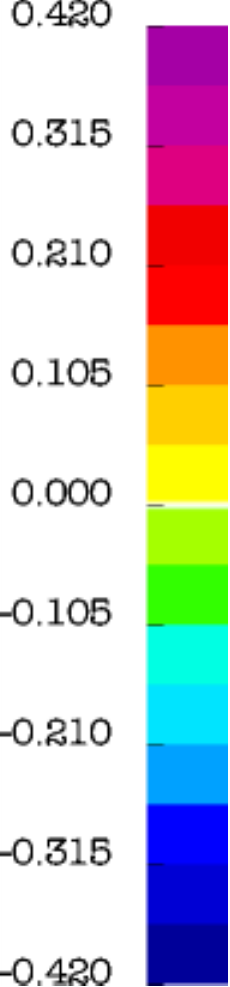}}\hss}
\end{center}
\caption{(Color online) A sub-Planck structure of the Wigner function evaluated for
  $\tau = 0.16$, $\tau = 0.3$, $\tau = 0.6$, $\tau = 1$.}
\label{Fig:10}
\end{figure*}

The sub-Planck structure is also present for the rational values of
evolution parameter, for which the state (\ref{eq:kerr}) does not
correspond to any coherent superposition state. It is depicted on
figure \ref{Fig:10} for $\tau=0.16$, $\tau=0.3$, $\tau=0.6$, $\tau=1$
and $\alpha=5$.  This structure is of different form.  The areas of
positive and negative values form ``ribbons.''

\section{Comparison of the Numerical Methods}\label{fos}

Figures \ref{Fig:1}--\ref{Fig:4}, presenting the Wigner function
evolution in an ideal $\chi^{(3)}$ medium, have been obtained by three
independent numerical simulations: computing of the equations
(\ref{eq:Fokker-Planck}), (\ref{summation}) and~(\ref{derivatives}).
The numerical results of the evolution in a noisy medium have been
achieved using only the Fokker-Planck equation (\ref{eq:Fokker-Planck}).

All programs used to simulate the Wigner function were written in
standard C++ language, due to its high speed and good portability.
The structure of the programs used for computation of the
equations~(\ref{summation}) and~(\ref{derivatives}) was based strictly
on these equations: they both consist of two loops which iterate on
$q$ and $k$ (or $m$ and $n$, respectively) and sum up the evaluated
coefficients for a given $\tau$ and $\gamma$.  Common terms were moved
out from the loops (especially the inner one) to optimize the
evaluation of the sums.  The main function was enclosed in
input/output logic which calls it for different values of $\gamma$ and
writes the results to a file.

The equations~(\ref{summation}) and~(\ref{derivatives}) involve
summation of the infinite series.  However, only a finite number of
their terms can be computed.  Therefore, assuming a given precision of
the simulation, we cut the series off.

The series in equation~(\ref{summation}) is much slower convergent
than the series in~(\ref{derivatives}) because it contains fast
oscillating exponential term $e^{-|\alpha|^2 e^{i\tau(k-q)}}$.
Depending on the phase factor in the exponent, this term can take
either large or small values, which makes impossible to compute
equation~(\ref{summation}) using the standard double precision
floating point number representation.  Therefore, for $|\alpha| \le 5$
both infinite sums, internal and external, are cut off by including at
least $500$ terms in each of them.  Also very high precision (at least
25 significant digits) needs to be applied for the computation.  Such
a high precision was obtained using the free {\it Class Library for
  Numbers} (CLN) and its long float type.

On contrary, it is enough to use standard double precision numbers to
compute equation~(\ref{derivatives}).  However, we also applied the
CLN here due to the fact that it can deal with the complex numbers
  while C++ itself cannot.  The same precision was used while
computing equation~(\ref{derivatives}) however, it was enough to take
$100$ terms in both sums into account to obtain the same results as
these which were achieved by computing~(\ref{summation}).  This made
simulations of equation~(\ref{derivatives}) much faster.
Additionally, the product of the exponens derivatives was replaced
with the following formula~\cite{bronstein}
\begingroup
\setlength{\arraycolsep}{0cm}
\begin{eqnarray}
(\partial_{\gamma})^n (\partial_{\gamma^*})^m e^{-4|\gamma|^2} &=&
  \sum_{k=0}^n {n\choose k} (\partial_{\gamma})^k (-4\gamma)^m
  (\partial_{\gamma})^{n-k} e^{-4|\gamma|^2}
\nonumber\\
&=& e^{-4|\gamma|^2} \sum_{k=0}^{\mathrm{min}(n,m)} {n\choose k} \frac{m!}{(m-k)!}
\nonumber\\
&\times& (-4\gamma)^{m-k} (-4\gamma^*)^{n-k}.
\end{eqnarray}
\endgroup

Both methods, equations~(\ref{summation}) and~(\ref{derivatives}),
allow for computation of the Wigner function for an arbitrary
evolution parameter $\tau$ and point $\gamma$ in the phase space.
Evaluation of the function for a chosen $\tau$ does not require
simulation of the whole evolution.  To visualize the Wigner function
we calculated it for each point of a $100\times 100$ square mesh of
the size proportional to the value of $\alpha$ ($-5\ldots 5$ for
$\alpha = 2$, $-8\ldots 8$ for $\alpha = 5$,
$-\frac{1}{2}\ldots\frac{1}{2}$ for sub-Planck structures), to show
all interesting aspects of the simulated function.

On the contrary, numerical simulation of the Fokker-Planck
equation~(\ref{eq:Fokker-Planck}) involves solving a partial
differential equation (PDE), which in turn requires setting a grid
(discretized phase space) and solving a set of linear equations (one
equation for each point of the grid) at each time step.

To solve the PDE, the equation~(\ref{eq:Fokker-Planck}) was rewritten
to difference quotient form, where all the partial derivatives were
replaced with appropriate difference quotients of fourth order to keep
the order of the quotients even and higher than the order of the
equation at the same time.  Even difference quotients are more
convenient, since they are symmetric with respect to the point where
the derivative is taken. The quotients were determined using the {\it
  Mathematica} program and its \verb|FiniteDifferencePolynomial|
function. We used polar mesh coordinates with constant radial distance
$\Delta r$ and constant angle $\Delta\varphi$ between points.  Below
we present the first, second and third derivatives discretized up to
the forth order:
\begingroup
\setlength{\arraycolsep}{0pt}
\begin{eqnarray}
  \partial_x & \rightarrow&
  \frac{f_{k - 2} - 8 f_{k - 1} + 8 f_{k + 1} - f_{k + 2}}{12 h}\\
  \partial^2_x & \rightarrow&
  \frac{- f_{k - 2} + 16 f_{k - 1} - 30 f_k + 16 f_{k + 1} - f_{k + 2}}{12 h^2}\\
  \partial^3_x & \rightarrow&
  \frac{-f_{k - 2} + 2 f_{k - 1} - 2 f_{k + 1} + f_{k + 2}}{2 h^3}
\end{eqnarray}
\endgroup
where $x=r,\varphi$, $h=\Delta r,\Delta\varphi$ and $k=i,j$ for $r$
and $\varphi$ respectively, $f_k$ is the Wigner function value.  The
mixed derivatives where built up combining the basic above formulas.

Since the standard {\it explicite} computation methods turned out to
be unstable due to large number of time steps, we applied {\it
  implicite} method. At the same time, it also allowed for larger time
steps. The main idea of {\it implicite} method is to determine the
time derivative between steps $\tau$ and $\tau + \Delta \tau$ using
the spatial derivatives obtained from the next time step
$W(\tau+\Delta\tau,\gamma,\gamma^*)$, not from the previous one
$W(\tau,\gamma,\gamma^*)$. This method requires solving a set of
linear equations, with Wigner function values in the next time step
$W(\tau+\Delta\tau,\gamma,\gamma^*)$ for each point $\gamma$ of the
mesh as unknowns.  The stability of this method relies on well-chosen
ratio of the time step and the grid density. For $\alpha=2$ we used
time step $\Delta\tau = \frac{\pi}{3600}$ and the grid consisting of
$300\;(\mathrm{radial \; direction})\times540\; (\mathrm{azimuthal\;
  direction)}$ points.  Thus, there were $162\cdot10^3$ points and
equations in the set.

In order to solve such a large set of equations, we put it in a matrix
form.  To achieve it we assigned an index $i$ for each point of the
polar mesh using the formula $i=N_f\,r/\Delta r +
\varphi/\Delta\varphi$, where $N_f$ is a density of the grid in
azimuthal direction, $r$ and $\varphi$ are polar coordinates of a
point of the mesh and $\Delta r$ and $\Delta\varphi$ are radial and
angular distances between mesh points.  This index was then used as a
coordinate to the row and column of the matrix where the coefficients
of the equations were placed and to rewrite the Wigner function values
of the previous time step from grid to vector form and the result from
vector back to mesh form.

\begin{figure*}[h!]
\begin{center}
\scalebox{0.6}{\includegraphics{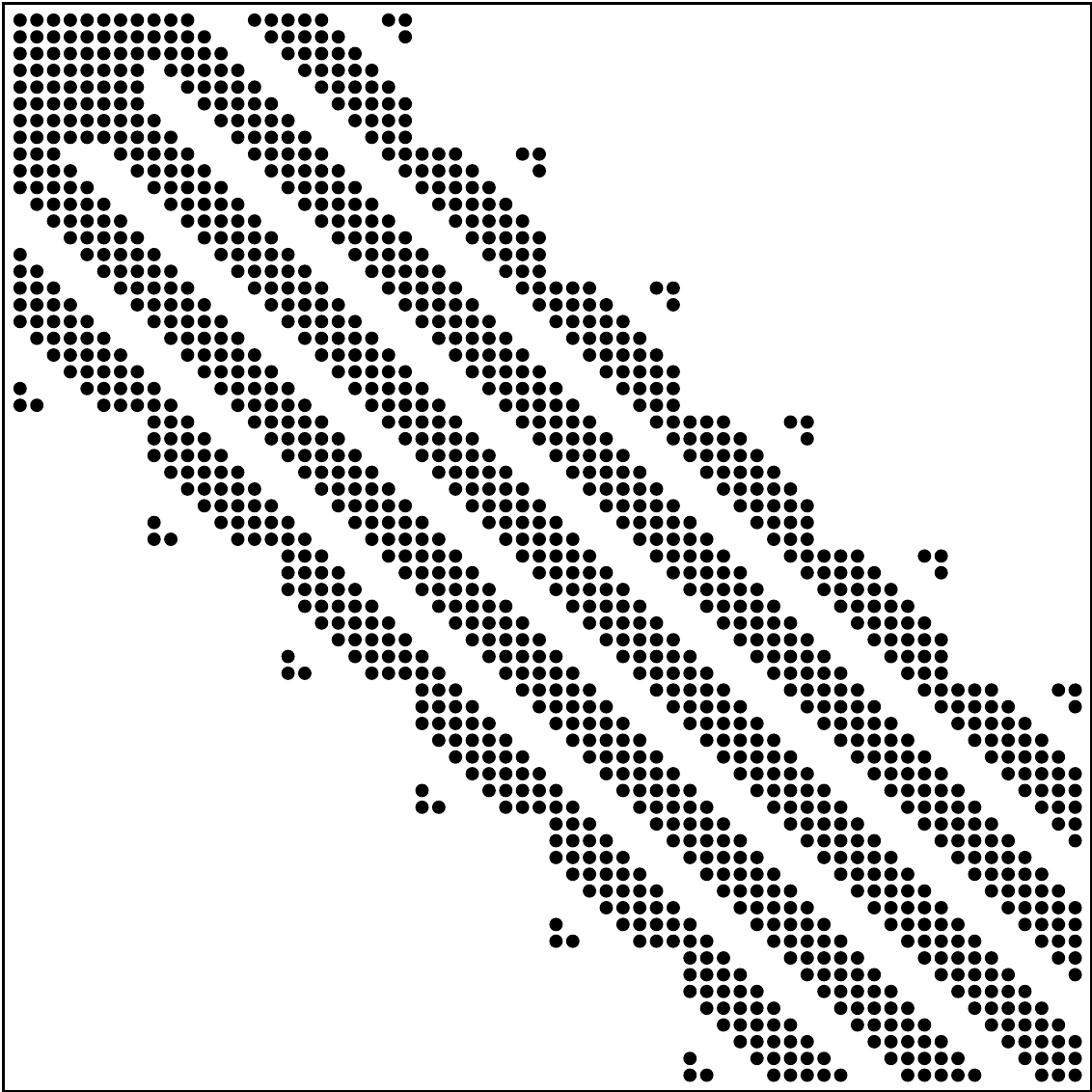}}
\end{center}
\caption{The sparse band matrix necessary to invert to solve the set
  of linear equations giving the Wigner function values for all points
  of the discretized phase space.}
\label{Fig:11}
\end{figure*}

The index $i$ (described above) was chosen in such a way that the
matrix of equation coefficients is a sparse band matrix.  The sample
of such matrix computed for small mesh size is depicted on
fig. \ref{Fig:11}.  Black points represent nonzero items (single
numbers).  It consists of five diagonal five-element bands and
additional two including either two or one element. In practice, the
matrix is much larger and has $162\cdot 10^3\times162\cdot 10^3$
elements, but the characteristic band diagonal structure with
five-element bands is preserved.

The set of equations is solved using \textit{Band Diagonal Systems}
method~\cite{nume}: the matrix is inverted applying LU decomposition
(Crout method) and then solved with Gaussian elimination while keeping
the matrices in special compressed form to save computer memory.  We
used the \verb|bandec| and \verb|banmul| routines from~\cite{nume},
translated to C++ and optimized.  We also used standard double
precision floating point numbers; there was no need for CLN library.
The main computing routine was accompanied by input/output logic used
to read simulation parameters and write out the results.

The flow of the program consisted of an introductory step --
performing a LU decomposition and calculating Wigner function for
$\tau=0$ (initial values for each point of the grid) -- and a sequence
of steps calculating evolution of the Wigner function for each
$\tau=\tau_0+n\Delta\tau$.  The solution gives the values of the
function for all $\gamma$ at each evolution step.

During the simulation some sanity checks were taken to make sure that
the calculations are done properly, e.g.\ the integral of the Wigner
function over the phase space is equal to 1 in each time step.  The
value of the integral unequal to 1 was a sign of either too small mesh
density or too big time step or too small precision of the
calculations.  We also compared the results of simulation of equations
(\ref{eq:Fokker-Planck}), (\ref{summation}) and~(\ref{derivatives})
for the same input parameters.

The numerical results are presented using OpenDX program.

\section{Conclusion}

In this paper we presented the Fokker-Planck equation which allowed
for numerical computation of the full Wigner function evolution
governed by the self-Kerr interaction Hamiltonian.  This equation can
be used for a state analysis for any input state, mixed or pure.  We
took a coherent state as an input state for the simulation.  For a
decoherence-free evolution the results have been obtained by three
different numerical algorithms.  We discussed the influence of the
decoherence process on the nonclassicality of the state under
evolution. For an exemplary calculation we took into consideration the
experimentally reasonable values of the nonlinearity and damping
constant for a nanomechanical resonator.  We also presented the
sub-Planck structure of the Wigner function which arises during the
evolution.  This equation can be applied to any system described by
the self-Kerr interaction.

This work was partially supported by a~MEN Grant No.~1 PO3B 137 30
(K.W.) and N202 021 32/0700 (M.S.).

\clearpage

\begin{thebibliography}{99}

\bibitem{Ludmila} L. Praxmeyer, P. Wasylczyk, C. Radzewicz, and
  K. W\'odkiewicz, \prl {\bf 98}, 063901 (2007).

\bibitem{Shang} S.-B. Li, X.-B. Zou, and G.-C. Guo, \pra {\bf 75},
  045801 (2007).


\bibitem{Agarwal} A. Biswas and G. S. Agarwal, \pra {\bf 75}, 032104
  (2007).

\bibitem{Grangier2007} A. Ourjoumtsev, A. Dantan, R. Tualle-Brouri,
  and Ph. Grangier, \prl {\bf 98}, 030502 (2007).

\bibitem{Grangier2006} A. Ourjoumtsev, R. Tualle-Brouri, and
  Ph. Grangier, \prl {\bf 96}, 213601 (2006).


\bibitem{Polzik} J. S. Neergaard-Nielsen, B. Melholt Nielsen,
  C. Hettich, K. Molmer, and E. S. Polzik, \prl {\bf 97}, 083604
  (2006).

\bibitem{Jeong} H. Jeong, A. P. Lund, and T. C. Ralph, \pra {\bf 72},
  013801 (2005).


\bibitem{Katz} I. Katz, A. Retzker, R. Straub, and R.  Lifshitz, \prl
  {\bf 99}, 040404 (2007).


\bibitem{Weetman} P. Weetman and M. S. Wartak, \prb {\bf 76}, 035332
  (2007).

\bibitem{Wigner} E. Wigner, Phys. Rev. {\bf 40}, 749 (1932).

\bibitem{Imoto} N. Imoto, H. A. Haus, and Y. Tamamoto \pra {\bf 32}, (1985).

\bibitem{Turchette} Q. A. Turchette, C. J. Hood, W. Lange, H. Mabuchi, and
H. J. Kimble \prl {\bf 75}, 4710 (1995).


\bibitem{Mohapatra} A. K. Mohapatra, M. G. Bason, B. Butscher, K. J. Weatherill,
    and C. S. Adams, quant-ph/0804.3273v1.


\bibitem{Imamoglu} A. Imamoglu, H. Schmidt, G. Woods, and M. Deutsch \prl {\bf 79}
1467 (1997). M. J. Werner and A. Imamoglu \pra {\bf 61} 011801 (1999).

\bibitem{Hau}  L. V. Hau, S. E. Harris, Z. Dutton, and C. H. Behroozi,
  Nature \textbf{397} 594 (1999).

\bibitem{Kang} H. Kang and Y. Zhu, \prl \textbf{91} 093601 (2003).

\bibitem{Bermel} P. Bermel, A. Rodriguez, J. D. Joannopoulos, and
    M. Soljacic, \prl \textbf{99}, 053601 (2007).


\bibitem{Brandao} F. G. S. L. Brandao, M. J. Hartmann, and M. B. Plenio,
    New J. Phys. \textbf{10} 043010 (2008).


\bibitem{Woolley} M. J. Woolley, G. J. Milburn, and Carlton
    M. Caves, quant-ph:0804.4540v1; E. Babourina-Brooks, A. Doherty,
    G. J. Milburn, quant-ph:0804.3618v1.


\bibitem{Kozinsky} I. Kozinsky, H. W. Ch. Postma, O. Kogan,
    A. Husain, and M. L. Roukes, \prl \textbf{99}, 207201 (2007).


\bibitem{Tanas} R. Tana\'s, {\em Nonclassical states of light
  propagating in Kerr media\/}, in Theory of Non-Classical States of
  Light, V. Dodonov and V. I. Man'ko eds., Taylor and Francis, London
  2003.

\bibitem{Walls} D. F. Walls, G. J. Milburn, \textit{Quantum Optics,}
  Springer-Verlag Berlin and Heidelberg GmbH and Co. KG (1995).



\bibitem{Milburn1986} G. J. Milburn and C. A. Holmes, \prl
  \textbf{56}, 2237 (1986).

\bibitem{Milburn1989} D. J. Daniel and G. J. Milburn, \pra
  \textbf{39}, 4628 (1989).

\bibitem{Perinova1990} V. Perinova and A.  Luks, \pra
  \textbf{41}, 414 (1990).

\bibitem{Gardiner} C. W. Gardiner, \textit{Quantum Noise},
  (Springer-Verlag, Berlin, 1991).




\bibitem{OSID} M. Stobi\'nska, G. J. Milburn, and K. W\'odkiewicz,
  OSID {\bf 14}, 81 (2007).

\bibitem{Averbukh1989} I. Sh. Averbukh and N. F. Perelman, Phys. Lett.
  \textbf{139}, 449 (1989).

\bibitem{Scully} M. O. Scully and M. S. Zubairy, \textit{Quantum
    Optics,} Cambridge University Press (1997).




\bibitem{Korsch1997} H. J. Korsch, C. Muller, and H. Wiescher, J.
  Phys. A {\bf 30}, L677 (1997).

\bibitem{Zurek} I. L. Chuang, R. Laflamme, P. W. Shor, W. H. Zurek,
  Science {\bf 270}, 1633 (1995).

\bibitem{ZurekN} F. Toscano, D. A. R. Dalvit, L. Davidovich, and
  W. H. Zurek, Nature {\bf 412}, 712 (2001).

\bibitem{ZurekPRA} W. H. Zurek, \pra {\bf 63}, 023803 (2006).


\bibitem{bronstein} I. N. Bronstein, K. A. Semendiajew, {\it Taschenbuch
  der Mathematik,} (B. G. Teubner Verlagsgesellshaft, Leipzig 1959).



\bibitem{nume} W. H. Pres, B. P. Flannery, S. A. Teukolsky, and
  W. T. Vetterling, {\it Numerical Recipes} (Cambridge University
  Press, Cambridge, 1988).



\end{thebibliography}
\end{document}